\documentclass[fleqn,usenatbib]{mnras}
\usepackage{newtxtext,newtxmath}
\usepackage[T1]{fontenc}
\DeclareRobustCommand{\VAN}[3]{#2}
\let\VANthebibliography\thebibliography
\def\thebibliography{\DeclareRobustCommand{\VAN}[3]{##3}\VANthebibliography}
\usepackage{graphicx}	
\usepackage{amsmath}	
\usepackage[nolist]{acronym}
\usepackage[noabbrev]{cleveref}
\usepackage[range-units = single]{siunitx}
\usepackage{booktabs}
\usepackage{bm}
\usepackage[table]{xcolor}
\usepackage{subcaption}
\usepackage{makecell}
\usepackage{tabularx}
\usepackage{multirow}

\newcommand\st{\textsf{\small ST}}
\newcommand\sts{\textsf{\small ST-S}}
\newcommand\stu{\textsf{\small ST-U}}
\newcommand\multinest{\textsc{MultiNest}}

\newcommand\SRGA{SRGA J144459.2$-$604207}

\begin{acronym}
\acro{1D}{one dimensional}
\acro{AMP}{accreting millisecond pulsar}
\acro{ART-XC}[ART-XC]{Astronomical Roentgen Telescope X-ray Concentrator}
\acro{cosi}[$\cos i$]{cosine of the observer inclination}
\acro{CDF}{cumulative distribution function}
\acro{CI}{credible interval}
\acro{d}[$D$]{distance}
\acro{GW}{gravitational waves}
\acro{GPU}{graphics processing unit}
\acro{EoS}{equation of state}
\acro{eXTP}[\textit{eXTP}]{\textit{enhanced X-ray Timing and Polarimetry}}
\acro{IXPE}[\textit{IXPE}]{\textit{Imaging X-ray Polarimetry Explorer}}
\acro{XMM-Newton}[\textit{XMM-Newton}]{\textit{X-ray Multi-Mirror Newton}}
\acro{i}[$i$]{observer inclination}
\acro{kch}{kilo core-hour}
\acro{KDE}{Kernel Density Estimation}
\acro{LMXB}{low-mass X-ray binary}
\acrodefplural{LMXB}{low-mass X-ray binaries}
\acro{m}[$M$]{mass}
\acro{MAP}{maximum a posteriori}
\acro{MP}{millisecond pulsar}
\acro{nh}[$N_{\rm H}$]{hydrogen column density}
\acro{NICER}[NICER]{Neutron star Interior Composition Explorer}
\acro{NS}{Neutron star}
\acro{obsid}{Observation ID}
\acro{phi}[$\phi$]{phase}
\acro{prior}{prior probability distribution}
\acro{posterior}{posterior probability distribution}
\acro{PPM}{pulse profile modelling}
\acro{PRE}{photospheric radius expansion}
\acro{r}[$R_{\rm eq}$]{equatorial radius}
\acro{rin}[$R_{\rm in}$]{inner disc radius}
\acro{RMP}{rotation powered millisecond pulsar}
\acro{SRG}[\textit{SRG}]{Spectrum-Roentgen-Gamma}
\acro{STU}[\textsf{ST-U}]{Single Temperature - Unshared}
\acro{tau}[$\tau$]{Thomson optical depth}
\acro{tbb}[$T_{\rm seed}$]{seed photon temperature}
\acro{TBO}{thermonuclear burst oscillation}
\acro{te}[$T_{\rm e}$]{electron temperature}
\acro{theta}[$\theta$]{co-latitude}
\acro{tc}[$\theta_{\rm c}$]{the constant parameters}
\acro{tin}[$T_{\rm in}$]{inner disc temperature}
\acro{X-PSI}[\textsc{X-PSI}]{\textsc{X-ray Pulse Simulation and Inference}}
\acro{zeta}[$\zeta$]{angular radius}
\end{acronym}

\newcommand\T{\rule{0pt}{2.8ex}}     
\newcommand\B{\rule[-1.6ex]{0pt}{0pt}} 

\title[PPM of SRGA J1444]{Pulse profile modelling of the 2024 outburst of the accreting millisecond pulsar SRGA J144459.2$-$604207}

\author[Dorsman and Salmi et al.]{%
Bas Dorsman,$^{1}$\footnotemark[1]
Tuomo Salmi,$^{2}$\thanks{Bas Dorsman and Tuomo Salmi are joint first authors with equal contribution. E-mail: b.dorsman@uva.nl and tuomo.salmi@helsinki.fi}
Anna L. Watts,$^{1,3}$
Mason Ng,$^{4,5}$
Anna Bobrikova,$^{6}$\newauthor
Alessandro Di Marco,$^{7}$
Duncan K. Galloway,$^{8,9}$
Sebastien Guillot,$^{10}$
Mariska Hoogkamer,$^{1}$
Yves Kini,$^{3}$ \newauthor
Fabio La Monaca,$^{7,11}$
Vladislav Loktev,$^{12}$
Matteo Lucchini,$^{1}$
Christian Malacaria,$^{13}$
Ying-Han Mao,$^{14,15}$ \newauthor
Alessandro Papitto$^{13},$
and Juri Poutanen$^{6}$
\\
$^{1}$Anton Pannekoek Institute for Astronomy, University of Amsterdam, Science Park 904, 1098XH Amsterdam, the Netherlands\\
$^{2}$Department of Physics, P.O. Box 64, FI-00014 University of Helsinki, Finland\\
$^{3}$Gravitation and Astroparticle Physics Amsterdam (GRAPPA), University of Amsterdam, Science Park 904, 1098XH Amsterdam, The Netherlands \\
$^{4}$Department of Physics, McGill University, 3600 rue University, Montr\'{e}al, QC H3A 2T8, Canada\\
$^{5}$Trottier Space Institute, McGill University, 3550 rue University, Montr\'{e}al, QC H3A 2A7, Canada\\
$^{6}$Department of Physics and Astronomy, 20014 University of Turku, Finland\\
$^{7}$INAF Istituto di Astrofisica e Planetologia Spaziali, Via del Fosso del Cavaliere 100, 00133 Roma, Italy\\
$^{8}$School of Physics and Astronomy, Monash University, Victoria 3800, Australia\\
$^{9}$Institute for Globally Distributed Open Research and Education (IGDORE)\\
$^{10}$University of Toulouse, CNES, CNRS, IRAP, Toulouse, France\\
$^{11}$Dipartimento di Fisica, Università degli Studi di Roma “Tor Vergata”, Via della Ricerca Scientifica 1, 00133 Roma, Italy\\
$^{12}$School of Mathematics, Statistics, and Physics, Newcastle University, Newcastle upon Tyne NE1 7RU, UK\\
$^{13}$INAF Osservatorio Astronomico di Roma, Via Frascati 33, 00078 Monte Porzio Catone, (RM), Italy\\
$^{14}$School of Astronomy and Space Science, Nanjing University, Nanjing 210023, P. R. China\\
$^{15}$Key Laboratory of Modern Astronomy and Astrophysics (Nanjing University), Ministry of Education, Nanjing 210023, P. R. China\\
}

\date{Accepted XXX. Received YYY; in original form ZZZ}
\pubyear{\the\year{}}

\begin{document}
\label{firstpage}
\pagerange{\pageref{firstpage}--\pageref{lastpage}}
\maketitle

\begin{abstract}
Pulse profile modelling via relativistic ray-tracing can constrain the system parameters of neutron stars, notably their mass and radius. Among these objects, accreting millisecond pulsars (AMPs) are promising targets, because they are bright in X-rays and their potentially polarized radiation can lead to complementary constraints on the emission geometry. We perform combined analysis of NICER and \textit{IXPE} observations of the recently discovered the 448-Hz pulsar SRGA~J144459.2$-$604207, with \textit{IXPE} providing X-ray polarization information. NICER and \textit{IXPE} jointly favour a large mass and radius for our best-fitting model, for which the neutron star has two independent hotspots. The primary hotspot is centered near the northern rotational pole, the secondary in the southern hemisphere, and the observer inclination is in the range 50–75 degrees. The primary hotspot is large (up to half the surface area) and contributes the majority of the non-pulsed X-rays, while the secondary is hotter and the major contributor to the overall pulse profile shape. However, many parameters are inferred to be near the prior bounds, which could indicate that the model does not adequately account for important physics. Furthermore, we tested several different methodologies for joint analysis of the two data sets: the results are sensitive to the method used, something that merits further study with synthetic data. In the future, we expect simultaneously recorded data will lead to improved parameter constraints, especially when multi-band and polarized data are combined.
\end{abstract}


\section{Introduction}
\acp{NS} are rich astrophysical laboratories for the study of the \ac{EoS} of ultra-dense nuclear matter, under conditions that cannot be replicated on Earth. Both the particle types (e.g. neutrons, hyperons, deconfined quarks) present in \ac{NS} cores, as well as their interactions, remain uncertain despite intense theoretical, experimental and astrophysical study \citep[see e.g.][]{Hebeler15,Tolos20,Lattimer21,Koehn25,Chatziioannou25}.

One way to constrain the \ac{EoS} is to measure the \ac{NS} \ac{m} and \ac{r}, since the \ac{m}-\ac{r} relation depends on the \ac{EoS} via the stellar structure equations. One can do this measurement using \ac{PPM} of millisecond X-ray pulsars, which are fast-rotating \acp{NS} with X-ray emitting hotspots. \Ac{PPM} is a relativistic ray-tracing technique used to estimate stellar parameters, such as \ac{m} and \ac{r}, by modelling the observed energy-phase dependent pulse profiles \citep[see e.g.][and references therein]{Pechenick1983,Miller1998,Poutanen2003,Morsink2007,Bogdanov2019b,Bogdanov2021}. The technique also constrains the properties of the hotspots, such as their size, location, and temperature.

So far, the technique has been most successfully applied to \acp{RMP}, for which the hotspots are thought to be generated at the magnetic poles due to heating by magnetospheric return currents \citep[see e.g.][]{Harding2001}. \Ac{PPM} has been applied to data from the \acl{NICER} (\acs{NICER}; \citealp{Gendreau2016})\acused{NICER}, often supplemented by data from the \ac{XMM-Newton} mission to help constrain the non-source background.  Current constraints on \ac{m} and \ac{r} are at the $\sim$$\pm$10~per cent level (68~per cent credible interval) and hotspot maps suggest that magnetic fields are not centred dipoles \citep{Riley2019, Bilous2019, Miller2019,Riley2021,Miller2021,Salmi2022,Salmi2023, Salmi2024a, Salmi2024b,Vinciguerra2024,Dittmann2024,Choudhury2024b,Qi2025,Hoogkamer25,Mauviard2025,Miller2025,Kini2026}. \acp{RMP} lend themselves well to \ac{PPM}, despite low X-ray fluxes, because their X-ray pulse profiles are stable over a long period of time, such that pulse profiles can be constructed from years of observational data. For several of these \acp{RMP}, radio timing provides informative priors on \ac{m},  improving the overall constraints \citep[see e.g.][]{Cromartie2020,Fonseca2021,Reardon2024}.

\ac{PPM} can also be applied to pulsations from rapidly-rotating accreting neutron stars in \acp{LMXB}. In such systems, when the low-mass companion fills its Roche lobe, matter is transferred through the inner Lagrange point, and owing to its specific angular momentum, the transferred material forms an accretion disc around the \ac{NS}. During outburst episodes \citep{Lasota2001,Heinke2025}, material from the inner edge of the disc is accreted onto the surface of the \ac{NS}, giving rise to X-ray emission. \Ac{PPM} can be applied to two types of pulsating \acp{NS} in \acp{LMXB}. The first type are \ac{TBO} sources. In that case, during Type I (thermonuclear) X-ray bursts, surface anisotropies on the rotating \ac{NS} appear, causing oscillations to be observed \citep{Watts2012,Bhattacharyya2022}. Performing \ac{PPM} for this class of pulsations has so far proved challenging due to pulse variability and computational cost \citep{Kini2023,Kini2024a,Kini2024b,Kini2025}. The second type are the persistent pulsations of \acp{AMP}, where the magnetic field of the \ac{NS} channels the accreted matter towards the magnetic poles of the \ac{NS}, leading to localized surface heating around the magnetic poles \citep{Patruno2021,DiSalvo2022}. For observers at favourable viewing angles, this leads to the detection of X-ray pulsations: 27 \acp{AMP} are now known. Due to Comptonisation of seed photons by hot electrons inside the in-falling material, the radiation should become linearly polarised \citep{Viironen2004}.

\acp{AMP} are interesting sources for PPM study. Firstly, because they are brighter (hundreds of counts per second, compared to less than 1 count per second for \acp{RMP}, see e.g. \citealt{Bult2019,Salmi2022}), they have the potential of providing tight constraints. Furthermore, due to their diverse phenomenology, other independent and complementary methods can be applied, permitting cross-verification of potential systematic errors in methods \citep{Watts16,extpdm}. Comparison to the results obtained for \acp{RMP} should also help us better understand the \ac{NS} population and formation history, because \acp{AMP} are believed to be the progenitors of \acp{RMP} \citep[although there are some caveats to this,][]{DAntona22}. \Ac{PPM} of \acp{AMP} can also place constraints on the accretion flow phenomena, such as the accretion disc and accretion column \citep{Poutanen2009,Ahlberg2024,Dorsman2025}.

However, \ac{PPM} of \acp{AMP} also introduces new challenges. \Acp{AMP} only sometimes go into outburst and their variability forces the selection of shorter data sets compared to \acp{RMP} (enough photons can typically still be gathered because \acp{AMP} are more luminous). There are also other sources of X-ray emission beyond just the surface hotspots. These include the inner disc region, as well as the accretion column. These components need to be taken into account in the \ac{PPM}, and may lead to degeneracies in the modelling \citep{Dorsman2025,Dorsman2026}. However, X-ray polarization data -- as recently provided by the \acl{IXPE} observatory (\acs{IXPE}; \citealp{IXPE2022,Papitto2025})\acused{IXPE} -- have been predicted to provide additional constraints that can alleviate this issue, because polarization provides an independent measure of the geometry \citep{Viironen2004,Salmi2021,Bobrikova2023,Salmi2025}. 

The first \ac{PPM} studies for \acp{AMP}, which laid much of the theoretical groundwork, made use of data from the \textit{Rossi X-ray Timing Explorer} \citep{Poutanen2003,Gierlinski2005,Leahy2008,Leahy2009, Morsink2011,Salmi2018}. More recently \citet{Dorsman2026} carried out \ac{PPM} using \ac{NICER} data for the \ac{AMP} SAX J1808.4$-$3658 from its 2019 and 2022 outbursts, finding a wide range of allowed parameters depending on how the accretion disc is modelled. They concluded that, in the soft X-ray band of \ac{NICER} (at least for this source), there exists a degeneracy between potential sources of the non-pulsed radiation component: radiation can come either from the disc or from the \ac{NS} surface, if part of the hotspot is continuously visible throughout a full rotation. Unfortunately, \ac{IXPE} was unable to observe SAX J1808.4$-$3658 during the 2022 outburst -- so there were no polarization data to break the degeneracy.\footnote{There are \ac{IXPE} observations of  SAX J1808.4$-$3658 from 2025 that are currently being analysed \citep{Ballocco2025}.} 

In February 2024, however, \citet{Molkov2024discovery} reported the discovery of a new 447.9-Hz \ac{AMP}, \SRGA. The \ac{AMP} nature was confirmed using NICER \citep{Ng2024}, and shortly afterwards \citet{Papitto2025} reported the discovery of polarized emission using data from \ac{IXPE}. This makes \SRGA{} the first AMP for which polarized pulse profiles are available. This opens up the possibility of doing joint \ac{PPM} for both data sets, in the hope of obtaining tighter constraints on the system parameters. This is the main goal of this paper.  

While previous \ac{PPM} studies have combined pulsed and unpulsed (phase-averaged) data sets, this is the first \ac{PPM} study to jointly analyse two pulsed data sets. Our secondary aim is thus to learn lessons for the future joint analysis of multiple data sets in the context of \ac{PPM} of \acp{AMP}. This topic is of major importance for the upcoming \ac{eXTP} mission \citep{extpdm,Zhang2025}, which has spectral and polarimetry instruments on board that can observe simultaneously.

This paper is structured as follows. \Cref{sec:method} describes the \ac{PPM} methodology applied here. \Cref{sec:dataprep} describes the preparation of the \ac{NICER} and \ac{IXPE} data. \Cref{sec:results} describes the results of the separate as well as joint \ac{PPM} analyses. \Cref{sec:discussion} discusses the results, and provides recommendations for future work. Finally, we conclude in \Cref{sec:conclusion}.

\section{Methodology}\label{sec:method}
To perform \ac{PPM}, we make use of the open source \acl{X-PSI} software (\acsu{X-PSI}, \texttt{v3.2.0}, \citealp{Riley2023}), used extensively for the analysis of \acp{RMP} and adapted for \ac{AMP} studies including polarimetric data \citep{Salmi2025,Dorsman2025,Dorsman2026}. We consider here two data sets, observed by \ac{NICER} and \ac{IXPE}, with the aim of improving the accuracy and precision of parameter inferences. For context, the two observations took place separated by a gap of 3.6 days (see \Cref{sec:dataprep} for more detail). The outburst flux and pulse profiles evolved between (and during) the observations. This variation implies that the parameters that govern the hotspots and accretion disc, for example, may change significantly between the two observations and we thus consider these `variable'. On the other hand a small subset of the parameters are considered `constant': \ac{m}, \ac{r}, \ac{d}, and \ac{cosi}, because they must be unchanging on the timescale of one outburst. We also initially assume \ac{nh} to be constant, but locally the effective X-ray absorption could evolve as the outburst proceeds, so we also consider analyses of the data where we assume this parameter to be variable. There are multiple Bayesian ways to analyse two sets of data, each with drawbacks and benefits, and here we will attempt three approaches. We will briefly introduce them now, but describe these in more detail in \Cref{sec:inference}. 

First, we analyse the two data sets independently. This approach is computationally feasible and a useful baseline, giving two independent sets of results that can be cross-verified. We also combine the independent \acp{posterior} after the fact, leading to joint \acp{posterior}. Second, we carry out a two-step Bayesian analysis where the \acp{posterior} from analysis of one data set are used as `updated' priors for the other (see \Cref{sec:inference} for a more detailed explanation). The prior update only takes place for any constant parameters; variable parameters are free to change and are thus not informed by a previous data set. This second strategy is also computationally affordable. In this paper we only update the prior density functions separately for each parameter using \ac{1D} cumulative distribution functions, noting that this method may result in a reduction of accuracy due to the loss of higher dimensional shape. Third, we construct a joint model to analyse the two data sets together, carrying out one nested sampling analysis with the joint model to infer parameters of both data sets simultaneously. Variable parameters will have two instances, one for each data set, meaning that the number of parameters governing this model is significantly increased. This method improves over the previous analysis in that it overcomes the loss of higher dimensional shape, but the increased number of parameters represents a considerable computational challenge. The remainder of this section describes the physical model deployed here, the parameter inference process, and finally, the \acp{prior} and the reasoning behind each.
 
\subsection{Physical model}\label{sec:model}
This section describes the model that produces the pulse profiles. This model is mostly the same as in \citet{Dorsman2025,Dorsman2026} and consists of two parts: the \ac{NS} and the accretion disc. Let us first describe the NS: the model that describes the X-ray emission of the surface hotspots. The underlying physics that governs the \ac{NS} surface emission has been studied over decades \citep{Pechenick1983, Riffert1988, Miller1998, Poutanen2003}. We give a short overview below, but for a comprehensive description we recommend \citet[][]{Bogdanov2019b}. The paths of light-rays that radiate off the \ac{NS} surface are computed in the oblate Schwarzschild + Doppler approximation, which assumes an oblate star embedded in a Schwarzschild spacetime \citep{Morsink2007, AlGendy2014}. Corrections are made to account for general relativistic light bending, light path time delay, emission angle aberration and relativistic Doppler boosting. This approximation is computationally efficient and sufficiently accurate for millisecond pulsars that rotate at a frequency below $\sim$600 Hz \citep[see e.g.][]{Oliva21,Silva2021,Jakab25}. The radiating hotspots on the \ac{NS} surface are parametrized by circles (although the hotspot shapes are three dimensional spheroidal caps). The co-latitudinal position is defined by $\theta_{\rm p/s}$\acused{theta} and phase $\phi_{\rm p/s}$\acused{phi} defines the azimuthal position of the centres of the primary/secondary hotspots. The angular radius of the hotspots is defined by $\zeta_{\rm p/s}$\acused{zeta}. Their atmospheric radiation pattern is given by the Comptonisation model of \citet{Bobrikova2023} which includes polarisation, and is governed by three parameters (omitting here the p/s subscript to simplify notation): \ac{tbb}, \ac{te}, and \ac{tau}. The atmospheric parameters are constant across the surface of the hotspots. We call the model of a \ac{NS} with a single hotspot of uniform distribution \st{} (for single temperature), \sts{} in the case of two hotspots that are antipodal twins with shared atmospheric parameters and \stu{} in the case of two independent (unshared) but non-overlapping hotspots. The hotspot emission bound for the observer is calculated at each rotational step of the \ac{NS} and collected in the appropriate phase and energy bin of the simulated pulse profile depending on energy shift and light path. 

The second part of the model is the accretion disc, which was first implemented into \ac{X-PSI} by \citet{Dorsman2025}. \citet{Dorsman2025} only implemented phase-independent disc emission (i.e. \texttt{diskbb}, see also \citealt{Mitsuda1984,Makishima1986}), but the presence of a disc naturally implies the geometric eclipse of \ac{NS} radiation by the disc. For this reason, we extend the model used in \citet{Dorsman2025, Dorsman2026} to include disc occultation. This model is used in the analysis of the \ac{NICER} data, but not in the final analysis of \ac{IXPE} data, as explained in \Cref{sec:results_ixpe}. This addition is also motivated by the findings of \citet{Molkov2024discovery}, who required partial hotspot occultation by the disc to reproduce the pulse profiles of \SRGA{} observed by the \ac{ART-XC} instrument on board the \textit{Spectrum-Roentgen-Gamma} observatory \citep{Sunyaev2021}.
To assess whether a photon trajectory is intercepted by the disc, we follow the approximate light-bending treatment in Appendix C of \citet{Ibragimov2009}. For each ray, we compute the radius at which it crosses the disc plane: if this radius is smaller than the \ac{rin}, the ray remains unobscured; otherwise, it is blocked. This typically affects mostly the secondary hotspot and does not yet include disc blocking of higher-order images, although that contribution is only relevant for very compact stars. This disc occultation module is implemented in \ac{X-PSI} from v3.2.0  (Mao, Dorsman et al. in prep). Afterwards, the non-eclipsed emission from the hotspots is added to the emission of the disc. The final pulse profile is obtained after attenuating the combined \ac{NS} and disc flux by interstellar absorption and convolving through the response of the observing telescope.  

\subsection{Parameter inference}\label{sec:inference}
The main goal of this work is to estimate the \acp{posterior} of model parameters. This section presents a short review of the Bayesian inference methodology relevant for this present work, followed by an description of the strategies explored to analyse the two data sets. The \ac{posterior} is expressed with Bayes' theorem as follows:

\begin{equation}\label{eq:posterior}
P(\theta|D)=\frac{\mathcal{L}(D|\theta)\pi(\theta)}{P(D)}.
\end{equation}

Here, $P(\theta|D)$, or \ac{posterior}, is the probability of the set of parameters $\theta$ in model $M(\theta)$ given data set $D$. The likelihood $\mathcal{L}(D|\theta)$ is the probability of obtaining data set $D$ given $\theta$. $\pi(\theta)$ is the prior, the probability of $\theta$ before considering $D$. The model $M(\theta)$ and parameters $\theta$ were described in \Cref{sec:model} and the \acp{prior} are described in \Cref{sec:priors}. An overview of the parameters and their \acp{prior} is given in \Cref{tab:parameters}. A common way to think about  Equation \ref{eq:posterior} is that it prescribes how prior knowledge of model parameters is `updated' upon the consideration of $D$, thus defining the \ac{posterior}.

$P(D)$, often denoted with the symbol $\mathcal{Z}$, is the probability of obtaining $D$, and is known as the marginal likelihood or evidence. Notice that $\mathcal{Z}$ does not depend on $\theta$, and therefore only serves as a normalization factor for the \ac{posterior}. When comparing two or more models however, $\mathcal{Z}$ is crucial because the ratio of two evidences defines the probability ratio of either model explaining the data, also known as the Bayes factor.

To estimate the \ac{posterior} we use nested sampling \citep{Skilling2004}. Nested sampling in Bayesian parameter inference is an algorithm that explores the parameter space more thoroughly and efficiently than ordinary random sampling. The top level idea is that the parameter space is mapped step by step using nested hypervolumes of prior space. The algorithm starts out by drawing samples from the full prior space but moves inward towards higher likelihood while recording an estimated prior probability `mass' that was removed at each step. Upon reaching some user-defined stopping criterion, one has obtained an estimate of the likelihood surface across parameter space, which is then used to estimate the \acp{posterior} and the overall $\mathcal{Z}$. We use the \multinest{} software \citep{Feroz2009,PyMultiNest} to perform this task. In our case, the stopping criterion is reached when the estimated possible remaining contribution from the current set of live points to the evidence $\Delta \ln(\mathcal{Z})$ is less than $0.5$ of the currently estimated evidence $\ln(\mathcal{Z})$ \citep[explained in more detail in][]{Feroz2009,Feroz2019}. This condition is no guarantee for convergence, and here we also test for convergence by increasing the number of live points until we find that posterior distributions no longer shift -- at least within the computational budget.

Now, we describe the strategies employed to separately and jointly analyse the two data sets. Analysis with a single instrument, as we will do with the individual \ac{NICER} and \ac{IXPE} data sets, is the most straightforward parameter inference scenario. The key ingredients are a function to sample parameters according to the prior, and a function that returns the ln-likelihood given some parameters and a data set. The case of \Ac{NICER} is simpler, for which the data set consists of a singular pulse profile consisting of X-ray counts. In case of \ac{IXPE}, we have 9 different data sets: Stokes $I$, $Q$, and $U$ parameters for 3 different detector units (DU$_{1,2,3}$). The total ln-likelihood is the sum over the likelihoods of the individual data sets. For the \ac{NICER} and \ac{IXPE} Stokes $I$ data, we use the Poisson likelihood because the X-ray counting is subject to Poisson noise. For \ac{IXPE} Stokes $Q$ and $U$, we use a Gaussian likelihood, which we discuss further in \ref{sec:ixpe_observations}. With these ingredients in hand, $P(\theta|D)\propto \mathcal{L}(D|\theta)\pi(\theta)$ can be estimated through the nested sampling process.

In this present work, we are also interested in doing a joint analysis of these two data sets. We describe here three different methods. \textbf{Combination method 1: combining independent posteriors}, is to estimate the joint \ac{posterior} by combining the results of the separate analyses of the \ac{IXPE} data ($D_{\rm I}$) and \ac{NICER} data ($D_{\rm N}$). To do so, we use a similar approach as the one presented in \citet[][see their section 2.2.4]{Kini2024b}. To take $D_{\rm N}$ as an example, the ln-likelihood surface is given by $\ln\mathcal{L}(D_{\rm N}|\theta_{\rm c},\theta_{\rm N}) = \ln P(\theta_{\rm c},\theta_{\rm N}|D_{\rm N})-\ln\pi(\theta_{\rm c},\theta_{\rm N})$, where we have separated \acsu{tc}, the subset of parameters that are constant between observations, and $\theta_{\rm N}$, the subset of variable parameters (defined in the beginning of \Cref{sec:method}), as observed by \ac{NICER}. In the case of \ac{IXPE}, these are $\theta_{\rm I}$.

We then approximate the joint ln-likelihood surface by adding both ln-likelihood surfaces. We do so by using \acp{KDE} of the posterior weighted samples and samples drawn from the priors. The joint \acp{posterior} are obtained by nested sampling again, now utilizing the \ac{KDE}-based joint ln-likelihood and the combined prior $\pi(\theta_c,\theta_{\rm I},\theta_{\rm N})$. The dimensionality of the final nested sampling step can be reduced by marginalizing out the subset of $\theta_{\rm I}$ and $\theta_{\rm N}$ that are only nuisance parameters (not affecting \ac{tc}). It turns out that only \ac{rin} is not a nuisance parameter. That is because the joint prior, $\pi(\theta_c, R_{\rm in})\not=\pi(\theta_c)\pi(R_{\rm in})$, because $\pi(R_{\rm in})=f(M, R_{\rm eq})$, thus causing prior correlation (see \Cref{sec:priors} for more detail on each prior). Still, the dimensionality of the joint sampling problem is greatly reduced to at most only 6 parameters: \ac{tc} (5 parameters) and $R_{\rm in,N}$. Because of the small dimensionality and the inexpensive ln-likelihood evaluation (from a \ac{KDE}) the final sampling step is computationally cheap and fast in real time.

\textbf{Combination method 2, two-stage Bayesian analysis}, is to do parameter inference in two steps. A first round of nested sampling estimates, for the sake of example, $\mathcal{D_{\rm N}}$: $P(\theta_{\rm c},\theta_{\rm N}|D_{\rm N})\propto \mathcal{L}(D_{\rm N}|\theta_{\rm c},\theta_{\rm N})\pi(\theta_{\rm c},\theta_{\rm N})$. We then define $\pi_{\rm updated}(\theta_{\rm c})=P(\theta_{\rm c}|D_{\rm N})$, and $\pi_{\rm updated}(\theta_{\rm c},\theta_{I})=\pi_{\rm updated}(\theta_c)\pi(\theta_{\rm I})$. In the case where $D_{\rm N}$ was analysed first, \ac{rin} is not a parameter in the subsequent \ac{IXPE} analysis, because no disc is used there. In the reverse case, if $D_{\rm I}$ was analysed first \ac{rin} exists and will be drawn using the updated prior distribution of \ac{m} and \ac{r}. In the \ac{NICER}-first case $P(\theta_{\rm c},\theta_{\rm I}|D_{\rm I},D_{\rm N})\propto\mathcal{L}(D_{\rm I}|\theta_{\rm c},\theta_{\rm I})\pi_{\rm updated}(\theta_{\rm c},\theta_{\rm I})$ is estimated during a second round of nested sampling that considers $D_{\rm I}$. The \acp{prior} of any parameters outside of \ac{tc} are not updated because they are variable and thus not informed by the other data set. In this example, $D_{\rm N}$ was analysed first, but the order of data set operations could be reversed and (in an idealized world) the analysis leads to equivalent results. However, in practice, a difference could occur due to insufficient exploration of the parameter space during sampling. On top of that, we  are only able to update  $\pi(\theta_{\rm c})$ with separate \ac{1D} \acp{prior}, leading to the loss of any higher dimensional correlation information.

\textbf{Combination method 3: simultaneous fit}, is to use a joint model that simultaneously fits two data sets $D_{\rm I}$ and $D_{\rm N}$, and fit parameters in one round of nested sampling. To do so, a joint \ac{posterior} is required, and it is given by
\begin{equation}
    P(\theta_c,\theta_I,\theta_N|D_I,D_N)\propto\mathcal{L}(D_{\rm I}, D_{\rm N}|\theta_c,\theta_{\rm I},\theta_{\rm N})\pi(\theta_c,\theta_{\rm I},\theta_{\rm N}).
\end{equation}
The observational data are independent of each other, so the joint likelihood factorizes as 
\begin{equation}\label{eq:jointlikelihood}
    \mathcal{L}(D_{\rm I}, D_{\rm N}|\theta_c,\theta_{\rm I},\theta_{\rm N}) = \mathcal{L}_{\rm I}(D_{\rm I}|\theta_c,\theta_{\rm I})\mathcal{L}_{\rm N}(D_{\rm N}|\theta_c,\theta_{\rm N}), 
\end{equation}
so in ln-space, the joint ln-likelihood is the sum of two ln-likelihoods: $\ln(\mathcal{L}) = \ln(\mathcal{L}_{\rm I})+\ln(\mathcal{L}_{\rm N})$, one for each data set.

In this method, there exist two instances of those parameters that can vary between the disjoint data sets, one for each data set. This enlarges the parameter space and makes the sampling problem harder. Because of this, and because the emission now needs to be calculated twice (one pulse profile for each data set), this run is more costly than methods 1 and 2. When we attempted to use this method, because of the computational expense, we could only use the limited (antipodal) \sts{} model, which led to worse fits. For completeness, we detailed the usage and results of this combination method in \Cref{sec:appendix_sts}. The main body of this work uses combination methods 1 and 2, which allowed the usage of the \stu{} model, leading to better evidence and better fits.
 
\subsection{Prior distributions}\label{sec:priors}
Following the discovery of \SRGA, a number of studies have used the data from the 2024 outburst to infer parameters, providing helpful prior information for this study. This section describes the prior distributions we adopt and \Cref{tab:parameters} provides an overview.

\begin{table}
    \caption{\label{tab:parameters} Overview of model parameters and their prior distributions. Some prior distributions are made up of two parts: the density distribution and bounds. In those cases the densities describe the shape of the prior, such as a normal distribution, but the cut-offs define hard lower and upper limits, denoted as (lower, upper). $U$(lower, upper) is a uniform distribution with lower and upper bounds. $N(\mu,\sigma)$ is a normal distribution with mean $\mu$ and a standard deviation $\sigma$. The subscripts ${\rm p}/{\rm s}$ indicate the primary and secondary hotspot, because these parameters exist for both hotspots. The bottom part of the table lists the $\alpha$-parameters, for which the priors change between the analyses of \ac{IXPE} data or the joint analysis (combination method 3).} 
    
    \begin{tabular}{lll} \hline \hline
    Parameter (unit) & Description & Prior density  \\ 
    \hline
    \multicolumn{3}{c}{Pulsar}\\
    \hline
    $M$ (M$_\odot$) & Mass & $U(1,2.2)$$^{a,b}$  \\
    $R_\mathrm{eq}$ (km) & Equatorial radius & $U(8,14)$$^{a,b}$  \\
    $D$ (kpc) & Distance & $N(8.5, 2)$$^c$   \\
    $\cos i$ (-)& Cosine inclination & $U(0, 0.64)$  \\
    $f$ (Hz) & Pulsar frequency & 447.871561   \\
    $N_\mathrm{H}$ (\SI{e22}{cm^{-2}}) & Column density & $U(1.9,2.9)$ \\
    $\chi_{0}$ (deg)& Spin axis position angle & $U(0, 360)$  \\
    \hline
    \multicolumn{3}{c}{Hotspots}\\
    \hline
    $\phi_{\rm p/s}$ (cycles) & Phase & $U(-0.25,0.75)$ \\
    $\cos\theta_{\rm p/s}$ & Cosine co-latitude & $U(-1, 1)$\\
    $\zeta_{\rm p/s}$ (deg) & Angular radius & $U(0, 90)$\\
    $T_{\rm seed,p/s}$ (keV) & Seed photon temperature & $U(0.5,1.5)$ \\
    $T_{\rm e,p/s}$ (keV) & Electron slab temperature & $U(20,100)$\\
    $\tau_{\rm p/s}$ (-) & Thomson optical depth & $U(0.5,3.5)$\\\hline
    
    \multicolumn{3}{c}{Disc}\\
    \hline
    $T_{\rm in}$ (keV) & Inner disc temperature & $U(0.01, 0.6)$\\
    $R_{\rm in}$ (km) & Inner disc radius & $U(R_{\rm eq},R_{\rm co})^b$ \\ \hline \hline
   \multicolumn{3}{c}{$\alpha$-parameters}\\
   \hline
    $\alpha_{\rm N}$ & Scaling of \ac{NICER} response & Fixed to 1$^d$  \\
                      & & $U(0.8,1.2)$ $^e$  \\
    $\alpha_{\rm DU1}$ & Scaling of \ac{IXPE} DU1 response &  $U(0.8,1.2)$$^d$  \\
         &  &  Fixed to 1 $^e$  \\
    $\alpha_{\rm DU2,DU3}$ & Scaling of \ac{IXPE} DU2/3 response &  $U(0.8,1.2)$ $^d$  \\
     &  & $U(0.95,1.05)$ $^e$  \\
    \hline
    \end{tabular}\\ 
    \vspace{2 mm}
    \begin{flushleft}
        \footnotesize{$^a$ These priors are also cut off by a compactness limit. See the text in \Cref{sec:priors} for more detail.}\\
        \footnotesize{$^b$ The priors of \ac{m}, \ac{r} and \ac{rin} depend on each other because samples are rejected if $R_{\rm in} > R_{\rm co} \propto M^{1/3}$ and $R_{\rm in} < R_{\rm eq}$. See the text in \Cref{sec:priors} for more detail.}\\
        \footnotesize{$^c$ This prior is truncated below 2.5kpc and above 10.6 kpc. }\\
        \footnotesize{$^d$ Individual analyses (when applicable), plus Combination methods 1 and 2.}\\
        \footnotesize{$^e$ Combination method 3.}\\
    \end{flushleft}
\end{table}

Here, we use a joint flat \ac{m}-\ac{r} \ac{prior}, which has been common in preceding \ac{PPM} analyses \citep{Riley2018}. However, the \ac{prior} was wide, typically $M$$\sim$$U(1,3)$ M$_\odot$ and \ac{r}$\sim$$U(3 R_{\rm g}(1),16)$, where $U$ is a flat distribution between a lower and upper limit, and $R_{\rm g}(1)$ is the gravitational radius ($GM/c^2$) of 1 M$_\odot$. While this choice of \ac{prior} is conservative, it produces many samples that are disfavoured by contemporary \ac{EoS} theory, experiments and astrophysical results \citep[see e.g.][]{Rutherford2024}. Instead, we use a joint flat prior with refined upper and lower limits derived from the \ac{EoS}-informed prior based on the piecewise-polytropic (PP) model described in \citet{Rutherford2024}. The prior for \ac{m} is $U(1,2.2)$M$_\odot$ and \ac{r} is $U(8,14)$ km. These bounds cover almost the whole prior distribution but exclude 1.4~per cent of the prior that has $M>2.2$M$_\odot$ and 2.3~per cent of the prior overall. We do not yet adopt the $M$-$R$ density distribution associated with \citet{Rutherford2024}, because at the time of the analysis it was not yet possible in \ac{X-PSI} to use multi-dimensional \acp{prior} of arbitrary shape. However, we do note that this would be the preferred method.

Furthermore, compactness limits were applied as usual. Initially, we reject samples beyond the causality limit given by $R_{\rm pole}/R_{\rm g}(M)>2.9$ \citep[see e.g.][]{Gandolfi2012}, where $R_{\rm pole}$ is the polar radius. However, for combination method 2, we increased the lower limit on radius of $R_{\rm pole}/R{\rm _g}(M)>3.01$.\footnote{We encountered numerical issues in \ac{X-PSI} when the stellar radius lies within its photon sphere (see \url{https://github.com/xpsi-group/xpsi/issues/633}). We therefore adopted the limit commonly used in \ac{NICER} \ac{RMP} analyses. This choice appears justified, as the final posteriors do not approach this regime.}

Constraints on the distance to the source come from the measurement of peak flux from the X-ray bursts observed with different instruments, coupled with assessment of the presence or absence of \ac{PRE} in the time-resolved spectroscopic measurements. \citet{Ng2024} and \citet{Malacaria2025Disk} derived upper limits of 10.6 and 11.5 kpc, respectively, based on the lack of \ac{PRE} inferred from the bursts observed with \ac{NICER} and {\it XMM-Newton}/{\it NuSTAR}. We adopt the strictest upper limit of the two at 10.6 kpc. We combine this the upper limit with the distance range estimated by \citet{Molkov2024discovery} of 8$-$9 kpc, based on more detailed spectral fits during bursts observed with {\it SRG}/ART-XC. Being conservative, we set up a normal distribution that is centred at the middle of this range, $\mu=8.5$, but is broader with $\sigma =$ 2 kpc, and a lower limit cut-off at $3\sigma$, corresponding to a lower limit of~2.5 kpc. We also note that, \citet{Fu2025comprehensive} estimate a distance of 10.03 $\pm$ 0.71 kpc, concluding that roughly 25\% of the bursts observed with {\it Insight-HMXT}\/ exhibited \ac{PRE}. However, we do not find the case for \ac{PRE} in those bursts convincing, as the time-resolved spectroscopic histories lack a significant radius maximum and contemporaneous temperature minimum \cite[see e.g.][]{Galloway2020}. 

For the observer inclination of the system, a breadth of values have so far been inferred. \citet{Papitto2025} fitted the observed pulsations by \ac{IXPE} to a simple \ac{NS} model. This was based on the analytical relations by \citet{Poutanen2006b} and \citet{Poutanen2020} with simplifying assumptions including a spherical NS (no oblateness), constant anisotropy of the emitted radiation and an energy-independent polarization. This model used fixed $M=1.4$M$_\odot$ and $R=10$ km, and point-like anti-phased hotspots. Using this model they fit Stokes $I$, $Q$ and $U$ and infer $i = 74.1^{+5.8}_{-6.3}\degr$. \citet{Malacaria2025Disk} perform spectral modelling on {\it XMM-Newton} and {\it NuSTAR} data and find $i=85^{+5}_{-14}\degr$ or $53\degr$ (depending on the approach taken to fit disc reflection). Finally, \citet{Mandal2025Relativistic} fit the persistent broadband spectra using \ac{NICER} and {\it NuSTAR} data and find $i = 50.3^{+2.0}_{-1.3}\degr$. We use a broad prior from 50\degr--90\degr, which is flat in \ac{cosi}, covering the breadth of the values that were inferred so far. 

We fix the pulsar spin frequency $f$ at 447.871561 Hz, as obtained by \citet{Ng2024}. For \ac{nh}, a range of values has been inferred, using different models. \citet{Ng2024} found (2.90 $\pm$ 0.03)$\times10^{22}{\rm\,cm^{-2}}$, based on a joint spectral fit of three \acp{obsid} of \ac{NICER} data. \citet{Malacaria2025Disk} found $(1.92^{+0.03}_{-0.01})\times10^{22}{\rm\,cm^{-2}}$ in their spectral fitting, and \citet{Mandal2025Relativistic} found $(2.5\pm0.1)\times10^{22}{\rm\,cm^{-2}}$. \citet{Li2025Timing} analyzed the X-ray broadband spectrum using \ac{NICER}, \textit{Einstein Probe}, \ac{IXPE}, \textit{Insight-HXMT}, and \textit{INTEGRAL} data. They found $(2.63 \pm 0.05)\times10^{22}{\rm\,cm^{-2}}$. We use a broad flat prior of $U(1.9,2.9)$$\times10^{22}{\rm\,cm^{-2}}$, which covers this range of values. 

Hotspot geometry: In the \stu{} configuration, the \ac{NS} features two independent hotspots (see Fig.~1 of \citealt{Vinciguerra2023}). The \acp{prior} related to hotspot position and size are kept broad: the prior distribution of \ac{phi} covers a full cycle: $U(-0.25,0.75)$, the \ac{theta} is flat in cosine and goes from north pole to south pole: $\cos\theta \sim U(-1,1)$, and the hotspot is allowed to have \ac{zeta} of up to half the star: $U(0,90)\degr$. However, the two hotspots are mutually constrained in their positioning and sizes in that they do not overlap and they must be ordered, meaning that the primary hotspot has a smaller co-latitude than the secondary. In practice, all samples that do not adhere to these requirements are rejected before costly computation. In the \sts{} configuration, the hotspots are identical and antipodal. That means that the number of hotspot parameters is halved because all parameters of the secondary spot are tied to the primary. In addition, $\cos \theta_{\rm p}$ (the primary hotspot) is constrained to the upper hemisphere $U(0,1)$, given that the \ac{NS} is north-south symmetric. 

Similar to \citet{Dorsman2025,Dorsman2026}, we do not adopt any prior information on the hotspot spectral parameters \ac{tbb}, \ac{te}, and \ac{tau}. Instead, we sample uniformly from the full table from \citet{Bobrikova2023}. For \ac{tin}, we use the same broad prior as \citet{Dorsman2026} from 0.01 to 0.6 keV. However, the shape is slightly changed because here we use a distribution flat in keV, whereas previously it was flat in $\log_{10}$(K). For \ac{rin} we also use the same prior as \citet{Dorsman2026}: a flat distribution between \ac{r} and the co-rotation radius $R_{\rm co}=(GM/4f^2\pi^2)^{1/3}$, where $G$ is the gravitational constant. As a result, the prior for \ac{rin} depends on other model parameters: \ac{r}, \ac{m} and the constant $f$. In practice, during sampling the value for \ac{rin} is drawn from within a 5--40~km  range, and samples that are outside the range  \ac{r}--$R_{\rm co}$ are rejected before any costly computation. Other works have found a breadth of values for $R_{\rm in}$. \citet{Molkov2024discovery} obtained 24.6 km based on fitting of the pulse profile that includes eclipses by the disc, but they did not fully explore the parameter space to obtain errors. $11\,R_{\rm g}\,(22.8\,M_{1.4}$ km, where $M_{1.4} = 1.4\,{\rm M}_\odot$) is found by \citet{Mandal2025Relativistic}, and $10.0^{+1.2}_{-1.7}\, R_{\rm g}\,(21_{-4}^{+3}\,M_{1.4}$km) or $6\,R_{\rm g}\,(12\,M_{1.4}$km) by \citet{Malacaria2025Disk}, where the latter is obtained if fixing $i=53\degr$. We note however that \ac{rin} is expected to be variable throughout the evolution of the outburst, and instruments have taken measurements at different times. We opt to stick with the broad prior described above.

Finally, we include $\alpha$-factors, scaling factors that are multiplied with the instrument responses to obtain the effective instrument responses, to account for potential deviation in cross-calibration. This is first done in the independent analysis of \ac{IXPE} data and combination method 2 (whenever \ac{IXPE} data is analysed) because \ac{IXPE} has three instruments on board. Here, the prior for $\alpha_{\rm DU1}$, $\alpha_{\rm DU2}$, and $\alpha_{\rm DU3}$ is $U(0.8, 1.2)$. Later, in combination method 3, where \ac{IXPE} and \ac{NICER} are jointly analysed, $\alpha$ are used again but are modified. There, we introduce $\alpha_{\rm N}$ with prior $U(0.8, 1.2)$, but we avoid increasing the parameter space by fixing $\alpha_{\rm DU1}$ to 1. While arbitrary, this restriction is a reasonable trade-off for computational efficiency, because it avoids an extra parameter that scales the flux, besides \ac{d}. To marginally shrink parameter space further, we also shrink the $\alpha_{\rm DU2}$ and $\alpha_{\rm DU3}$ \acp{prior} to $U(0.95, 1.05)$. This additional constraint is justified, because the \ac{IXPE}-only analysis resulted in highly correlated $\alpha$-factors (\Cref{fig:cornerplot_IXPEonly_rest}), meaning $\frac{\Delta \alpha_{\rm DU2}}{\Delta \alpha_{\rm DU1}} \approx \frac{\Delta \alpha_{\rm DU3}}{\Delta \alpha_{\rm DU1}} \approx 1.0$.

\section{Data preparation}\label{sec:dataprep}
\begin{figure}
    \centering    \includegraphics[width=\linewidth]{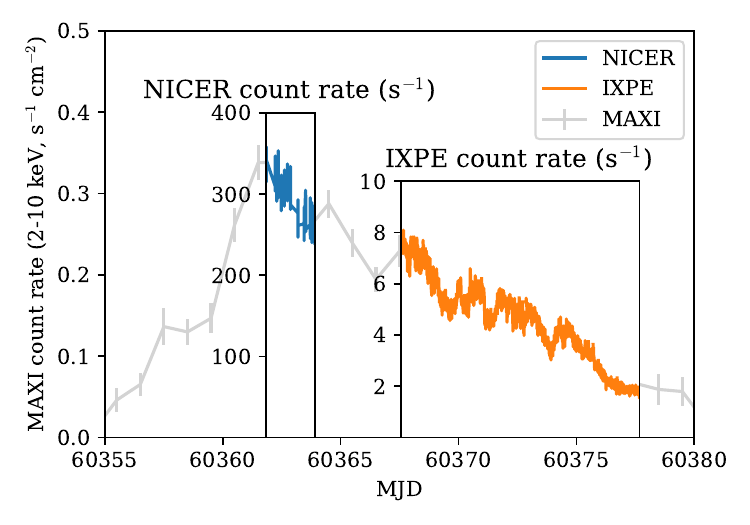}
    \caption{The outburst as captured by \textit{MAXI}, \ac{IXPE}, and \ac{NICER}. Only the \ac{NICER} data selected for the analysis are shown.}
    \label{fig:outburst}
\end{figure}

\begin{figure*}
    \centering
    \includegraphics[width=\linewidth]{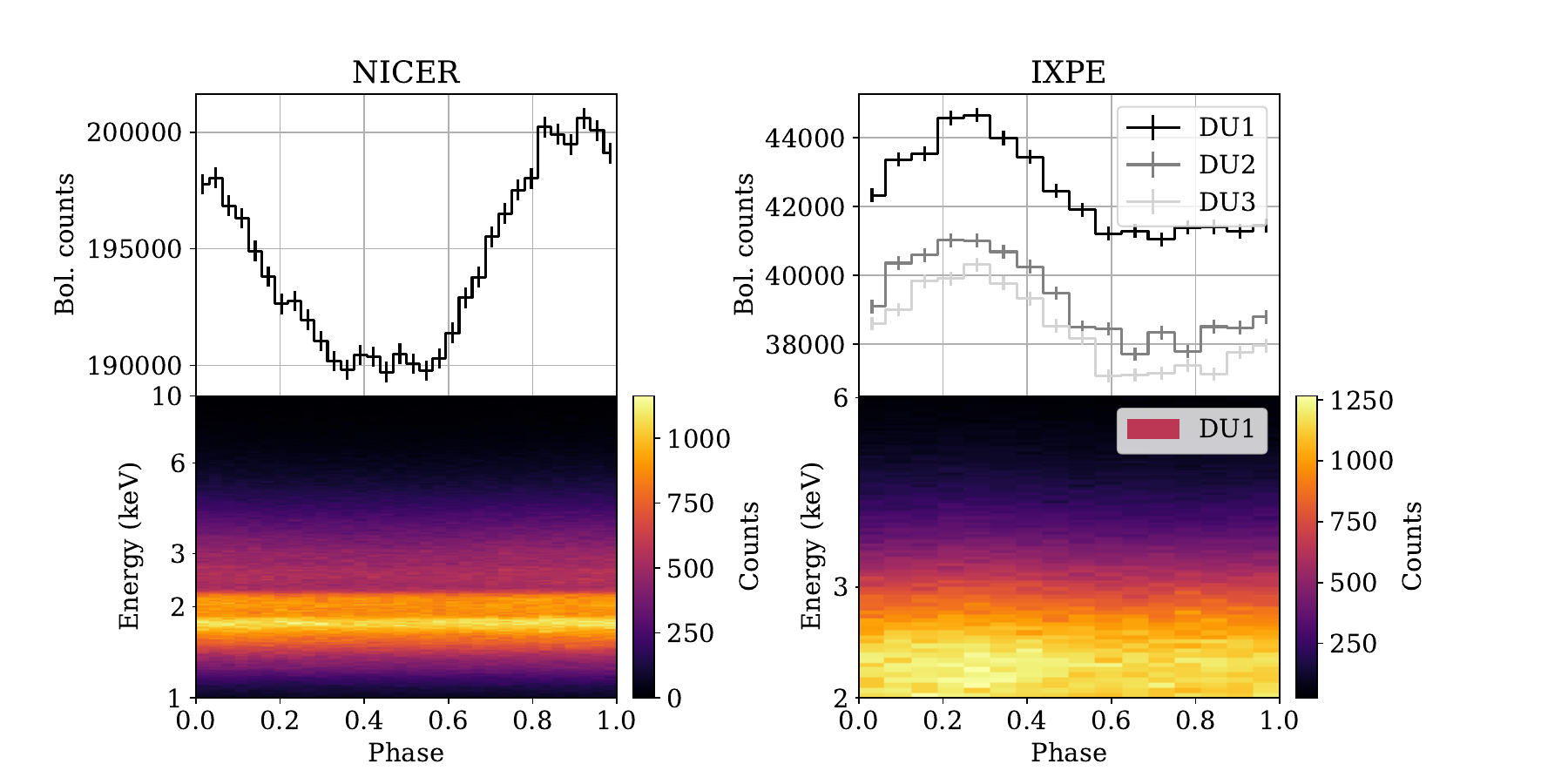}
    \caption{Pulse profiles of \ac{NICER} (left) and the three \ac{IXPE} detector units (right). Top panels: bolometric counts. Bottom panels: phase-energy resolved pulse profiles. For \ac{IXPE}, only the phase-energy resolved pulse profile of DU$_1$ is shown, because the other two are very similar.}
    \label{fig:pulseprofiles}
\end{figure*}

This section describes the preparation of the \ac{NICER} and \ac{IXPE} data into pulse profiles. \Cref{fig:outburst} shows the outburst as observed by \textit{MAXI} \citep{Matsuoka2009}, \ac{NICER}, and \ac{IXPE}. Only the subset of \ac{NICER} data selected for analysis is shown. \Cref{fig:pulseprofiles} shows the pulse profiles resulting from the data preparation process.   

\subsection{NICER Observations}\label{sec:dataprep_nicer}
Here, we outline the data processing steps undertaken for the \ac{NICER} observations used in the PPM analysis of \SRGA.  These are largely similar to those presented in \citet{Dorsman2026}, which analysed SAX~J1808.4$-$3658. \ac{NICER} is well-suited for the PPM analysis given its large effective area ($\sim$$1900{\rm\,cm^{2}}$ at 1~keV) and exquisite time-tagging capabilities in the 0.2--12.0~keV energy range \citep[100~ns absolute timing accuracy,][]{Gendreau2016,Prigozhin2016}. \ac{NICER} is an external payload on the International Space Station comprising 56 co-aligned X-ray concentrator optics and silicon drift detectors in focal plane modules (FPMs), of which 52 are operational \citep{LaMarr2016}.

For this work, we utilized the \ac{NICER} observations taken between 2024 February 21 to 2024 March 14, with \acp{obsid} 6204190101 to 6204190103 and 6639080101 to 6639080113 inclusive. However, we note that the observations in \acp{obsid} 6639080105 to 6639080112 suffered from large undershoot rates of over 700\,c\,s$^{-1}$ per FPM, most likely due to high levels of optical loading,\footnote{\url{https://heasarc.gsfc.nasa.gov/docs/nicer/analysis_threads/undershoot-intro/}} and thus did not provide any scientifically useful data (i.e. filtered exposure time of 0~s). The remaining data have very low undershoot rates of $\sim$3\,c\,s$^{-1}$ per FPM.

We processed the data with the \ac{NICER} Data Analysis Software (\textsc{NICERDAS}) v13 in \textsc{HEASoft} v6.34 with the \ac{NICER} calibration database (CALDB) version \textsc{xti20240206}. To generate the good time intervals (GTIs) for the \ac{PPM} analysis, we adopted the following filtering criteria: Earth limb elevation angle of ${\rm ELV} > 20\degr$; bright Earth limb angle of ${\rm BR\_EARTH > 30\degr}$; angular pointing offset (to the source) of ${\rm ANG\_DIST < 54}\arcsec$; an undershoot rate per FPM of 0--500~c\,s$^{-1}$; an overshoot rate per FPM of 0--30~c\,s$^{-1}$, and a magnetic cutoff rigidity of $>1.5{\rm\,GeV/c}$. Additionally, we visually identified and manually excised five time intervals (a total of 349~s) that contained type-I X-ray bursts (for more on the bursts, see \citealt{Molkov2024discovery}, \citealt{Fu2025comprehensive}, and \citealt{Papitto2025}). Finally, we have 24.8~ks of filtered exposure for our analysis.

To generate the pulse profiles for our PPM efforts, we first constructed a local timing solution by starting from the solution given in \citet{Papitto2025} and utilized PINT \citep{Luo2021} to model any remaining timing residuals. We derived a timing solution that is consistent (to within 1--$2\sigma$) of the published timing solution \citep{Papitto2025}. The selected \ac{NICER} data are recorded during the peak and early decrease of the outburst, as shown in \Cref{fig:outburst}. During this time, the pulse phase is stable, consistent with $\dot f=0$, similar to  the finding of \citet{Papitto2025}. For the construction of the pulse profile, any evolution of flux and pulse amplitude are averaged out. However, we note that the amplitude of the first harmonic (the fundamental frequency) appears to rise slightly from $2.4 \pm 0.4$ to to $3.3 \pm 0.4$~per cent, while the flux decreases from $\sim$ 350 to 250 c\,s$^{-1}$. 

Similarly to \cite{Dorsman2026}, we assessed whether constructing an averaged representative ancillary response file (ARF) or redistribution matrix file (RMF) would suffice for our work, or whether having observation-specific ARFs/RMFs would be more appropriate. We constructed the ARFs/RMFs for the individual observations by utilizing the \textsc{nicerarf} and \textsc{nicerrmf} tools\footnote{\url{https://heasarc.gsfc.nasa.gov/docs/nicer/analysis_threads/arf-rmf/}} (detectors 14 and 34 disabled). For the ARFs (i.e. effective area curves), the largest fractional difference between the observations used in this work was roughly 0.2~per cent. For the RMFs, there was virtually no difference\footnote{50 bins out of 6 million bins in element number-energy space showed discrepancies.} between the different observations. Thus it is appropriate to sum the ARFs and RMFs together into an averaged ARF/RMF for our analysis. The operation was facilitated by the \textsc{HEASoft} tasks, \texttt{ftaddrmf} and \texttt{ftaddarf}, where we constructed the exposure time-weighted ARF and RMF. 

We inspected the environmental and instrumental background spectrum, estimated by the 3C50  model \citep{Remillard2022}, in the energy range 0.3--10~keV and found that it varied from 0.5 to 4.5~per cent throughout the observation. This fraction is acceptably small, and these estimated background counts are included in the pulse profile data but not separately accounted for during \ac{PPM}, similarly to \citet{Dorsman2026}. Finally, very few counts, ${\sim}$1~per cent, are detected below channel 100 (${\sim}$1~keV). To save on computational effort calculating low energy radiation, we cut out the low energy channels below channel 100.

\subsection{IXPE Observations}\label{sec:ixpe_observations}
For this work we utilize also the \ac{IXPE} data, extracted from the same observation period as in \citet{Papitto2025}, i.e. starting shortly after the \ac{NICER} observation finished (IXPE GO 03250101; PI: A.~Papitto). \ac{IXPE} is well-suited for spectro-polarimetric timing studies owing to its imaging X-ray polarimetry capabilities in the 2--8 keV energy band, combining three co-aligned mirror modules with a net effective area of $\sim$$100~{\rm cm^2}$ and event-by-event measurements providing spatial, spectral, timing, and polarization information, with $\lesssim 10~\mu$s timing resolution \citep{Soffitta2021,IXPE2022}.

In contrast to the approach chosen for the \ac{NICER} data, we instead adopted here the same timing solution as \citet{Papitto2025} to construct the pulse profiles. The observation lasted $\sim$10~d and we used the \texttt{xpbin} tool in \textsc{ixpeobssim} package version 30.6.4 \citep{Baldini2022} to extract weighted $I$, $Q$ and $U$ spectra \citep{DiMarco_2022} along with the \ac{IXPE} CALDB \textsc{v013} of January 2024 \citep{Kislat2015}. 
Source photons are extracted from a circular region with a radius of 100\arcsec\ centred at the source position; considering the source flux, background rejection or subtraction is not needed \citep{DiMarco_2023}. The model-independent \texttt{pcube} algorithm was employed to determine the polarization quantities as a function of pulse phase in \citet{Papitto2025}. Here, we instead performed a spectropolarimetric analysis in 16 phase bins. For the \ac{PPM} analysis we selected only phase-energy bins with nominal energies in the 2--6~keV range, where the number of $I$ counts in each bin is $\gtrsim 20$ without additional rebinning. In this regime, the Stokes parameters $Q$ and $U$ are approximately normally distributed, and we therefore model their likelihoods with Gaussian functions using the measured uncertainties as standard deviations  \citep[see][for details]{Baldini2022}.\footnote{For fewer counts per bin, a Gaussian function is not a good approximation for the distribution of counts. A Poisson function is also not valid due to the presence of negative counts. However, Gaussian functions could still be used after re-binning or excluding nearly empty bins.}

\section{Results}\label{sec:results}
This section presents the results from the \ac{PPM} analyses of the \ac{NICER} and \ac{IXPE} data. First, the data sets are analysed individually in \Cref{sec:results_nicer,sec:results_ixpe}. Next, we combine analysis of both data sets in multiple ways. In \Cref{sec:results_kde}, we combine the likelihood surfaces of the separate analyses with combination method 1, as described in \Cref{sec:method}. Following this, in \Cref{sec:results_posterior_informed}, we apply combination method 2. \Cref{tab:inferences} gives an overview of the inferred parameter values that result from these methods, which are the main results of this work. However, in \Cref{sec:appendix_sts}, we explored also the usage of combination method 3. The results of that are not included in the main body of the paper because only the less preferred \sts{} model could be used. An overview of those results is given in \Cref{tab:inferences_appendix}. These two aforementioned tables specify also relevant sampling parameters: number of parameters, the nested sampling efficiency, run cost, and number of live points. The text related to each run discusses the choice in live points and convergence.

\begin{table*}
\centering
\caption{An overview of the analyses done in this work and the inferred parameters. \label{tab:inferences} In the header, various analysis and fit statistics are shown. The sampling efficiency is the nested sampling efficiency as defined in \ac{X-PSI}. This number is modified before being passed to \multinest, because \multinest{} is unaware some samples are rejected in the \ac{prior}. $\chi^2$ is calculated using 100 draws from the posterior, as described in the text. The $\hat{\chi}^2=\chi^2$/DOF, and the p-value are given based on the X-ray counts without taking into account polarization. For the \ac{IXPE} $I$, $Q$, and $U$ runs, the $\chi^2$/DOF and the p-value are given for the combined Stokes $I$ data of the DU$_1$, DU$_2$ and DU$_3$ detectors. For the KDE combination, the sampling parameters of the last sampling step are given. The body of the table gives an overview of the posterior median values and 68~per cent \acp{CI} of all parameters.} 

\begin{tabular}{lccccccc}
\hline
\hline
Name & \textbf{NICER-only} & \textbf{\ac{IXPE}-only ($IQU$)} & \textbf{\ac{IXPE}-only ($QU$)}  & \makecell{\textbf{Combination} \\ \textbf{method 1}} & \multicolumn{3}{c}{\textbf{Combination method 2}} \\
\hline

Data & NICER & \ac{IXPE} $IQU$ & \ac{IXPE} $QU$ & \makecell{NICER,\\ \ac{IXPE} $IQU$} & NICER & NICER & \ac{IXPE} $IQU$ \T\B\\
Prior updated by & - & - & - & - & \ac{IXPE} $IQU$ & \makecell{\ac{IXPE} $IQU$, \\ flat \ac{nh}} & NICER \T\B\\
Hotspot config. & \stu & \stu & \stu & - & \stu & \stu & \stu \\
\textbf{Sampling aspects} &  &  &  &  &  &  & \T\B\\
Num. params. & 19 & 21 & 21 & 5 & 19 & 19 & 21 \\
Samp. eff. (X-PSI) & 0.1 & 0.1 & 0.1 & 0.1 & 0.1 & 0.1 & 0.3 \\
Run cost (kch) & 2.9 & 45.8 & 0.4 & $\lesssim0.1$ & 9.3 & 18.5 & 13.3 \\
Live points & 2000 & 4000 & 10000 & 4000 & 2000 & 2000 & 4000 \\
$\ln(\mathcal{Z})$ 
& $-103760$ 
& $-55372$ 
& $-43479$ 
& $-22.792$ 
& $-103800$ 
& $-103740$ 
& $-55369$ \\
\textbf{Fit quality} &  &  &  &  &  &  & \T\B\\
$\chi^2$/DOF & 30108/28781 & 4940/4827 & - & - & 30204/28781 & 30103/28781 & 4953/4827 \\
$\hat{\chi}^2$ & 1.0461 & 1.0233 & - & - & 1.0494 & 1.0459 & 1.0262 \\
p-value & $2.4792\times 10^{-8}$ & $1.2632\times10^{-1}$ & - & - & $2.6326\times10^{-9}$ & $2.8227\times10^{-8}$ & $9.9836\times10^{-2}$ \\

\hline
\rowcolor{gray!15}$M\;\mathrm{[M}_{\odot}\mathrm{]}$ 
& $2.17_{-0.03}^{+0.02}$ 
& $2.02_{-0.36}^{+0.12}$ 
& $1.4_{-0.3}^{+0.4}$ 
& $1.928^{+0.012}_{-0.012}$ 
& $2.15_{-0.06}^{+0.03}$ 
& $2.15_{-0.05}^{+0.03}$ 
& $2.16_{-0.03}^{+0.02}$ \T\B \\

$R_{\mathrm{eq}}\;\mathrm{[km]}$ 
& $13.82_{-0.14}^{+0.10}$ 
& $10.7_{-1.0}^{+2.4}$ 
& $13.1_{-0.8}^{+0.6}$ 
& $13.94^{+0.04}_{-0.05}$ 
& $11.1_{-0.3}^{+0.3}$ 
& $13.6_{-0.3}^{+0.3}$ 
& $13.85_{-0.11}^{+0.09}$ \T\B \\

\rowcolor{gray!15}$D \;\mathrm{[kpc]}$ 
& $2.88_{-0.04}^{+0.05}$ 
& $3.3_{-0.3}^{+0.3}$ 
& $3.3_{-0.5}^{+0.8}$ 
& $3.060^{+0.017}_{-0.019}$ 
& $2.71_{-0.07}^{+0.07}$ 
& $2.55_{-0.03}^{+0.05}$ 
& $2.88_{-0.03}^{+0.04}$ \T\B \\

$i\;\mathrm{[deg]}$ 
& $50.9_{-0.5}^{+0.8}$ 
& $63_{-8}^{+12}$ 
& $77_{-10}^{+8}$ 
& $54.2^{+0.2}_{-0.3}$ 
& $53_{-2}^{+3}$ 
& $51.1_{-0.8}^{+1.4}$ 
& $50.8_{-0.5}^{+0.7}$ \T\B \\

\rowcolor{gray!15}$N_\mathrm{H}\;[10^{21} \mathrm{cm}^{-2}]$ 
& $27.82_{-0.08}^{+0.07}$ 
& $22.1_{-0.9}^{+1.0}$ 
& $24_{-3}^{+3}$ 
& $27.59^{+0.03}_{-0.03}$ 
& $26.340_{-0.015}^{+0.008}$ 
& $28.04_{-0.11}^{+0.10}$ 
& $27.82_{-0.08}^{+0.07}$ \T\B \\

$\phi_\mathrm{p}\;\mathrm{[cycles]}$ 
& $-0.34_{-0.04}^{+0.05}$ 
& $-0.06_{-0.17}^{+0.14}$ 
& $0.28_{-0.70}^{+0.16}$ 
& - 
& $-0.45_{-0.02}^{+0.02}$ 
& $-0.32_{-0.07}^{+0.27}$ 
& $-0.25_{-0.03}^{+0.05}$ \T\B \\

\rowcolor{gray!15}$\theta_\mathrm{p}\;\mathrm{[deg]}$ 
& $1.1_{-0.2}^{+0.3}$ 
& $0.7_{-0.3}^{+0.4}$ 
& $17_{-9}^{+50}$ 
& - 
& $3.0_{-0.9}^{+1.4}$ 
& $1.1_{-0.3}^{+0.3}$ 
& $1.5_{-0.3}^{+0.4}$ \T\B \\

$\zeta_\mathrm{p}\;\mathrm{[deg]}$ 
& $88.6_{-1.2}^{+0.9}$ 
& $82_{-10}^{+5}$ 
& $43_{-29}^{+15}$ 
& - 
& $86_{-3}^{+3}$ 
& $86_{-3}^{+3}$ 
& $83_{-7}^{+5}$ \T\B \\

\rowcolor{gray!15}$T_\mathrm{seed,p}\;\mathrm{[keV]}$ 
& $0.822_{-0.006}^{+0.009}$ 
& $0.85_{-0.14}^{+0.05}$ 
& $1.39_{-0.60}^{+0.11}$ 
& - 
& $0.97_{-0.03}^{+0.03}$ 
& $0.795_{-0.013}^{+0.014}$ 
& $0.649_{-0.009}^{+0.008}$ \T\B \\

$T_\mathrm{e,p}\;\mathrm{[keV]}$ 
& $25.9_{-0.5}^{+0.6}$ 
& $76_{-13}^{+15}$ 
& $60_{-20}^{+20}$ 
& - 
& $23.0_{-0.8}^{+0.9}$ 
& $24.2_{-0.8}^{+0.8}$ 
& $40_{-3}^{+4}$ \T\B \\

\rowcolor{gray!15}$\tau_\mathrm{p}\;[-]$ 
& $2.003_{-0.023}^{+0.019}$ 
& $1.18_{-0.07}^{+0.06}$ 
& $2.0_{-0.7}^{+0.8}$ 
& - 
& $2.15_{-0.04}^{+0.04}$ 
& $2.09_{-0.03}^{+0.03}$ 
& $1.39_{-0.05}^{+0.05}$ \T\B \\

$\phi_\mathrm{s}\;\mathrm{[cycles]}$ 
& $-0.420_{-0.006}^{+0.006}$ 
& $-0.311_{-0.012}^{+0.049}$ 
& $-0.1_{-0.2}^{+0.3}$ 
& - 
& $-0.466_{-0.014}^{+0.016}$ 
& $-0.418_{-0.009}^{+0.016}$ 
& $-0.286_{-0.012}^{+0.010}$ \T\B \\

\rowcolor{gray!15}$\theta_\mathrm{s}\;\mathrm{[deg]}$ 
& $145_{-3}^{+3}$ 
& $146_{-10}^{+13}$ 
& $156_{-50}^{+17}$ 
& - 
& $143_{-5}^{+5}$ 
& $146_{-4}^{+5}$ 
& $139_{-13}^{+8}$ \T\B \\

$\zeta_\mathrm{s}\;\mathrm{[deg]}$ 
& $52_{-3}^{+3}$ 
& $22_{-5}^{+8}$ 
& $51_{-36}^{+18}$ 
& - 
& $59_{-6}^{+7}$ 
& $51_{-28}^{+7}$ 
& $13_{-4}^{+9}$ \T\B \\

\rowcolor{gray!15}$T_\mathrm{seed,s}\;\mathrm{[keV]}$ 
& $0.84_{-0.04}^{+0.04}$ 
& $1.34_{-0.16}^{+0.12}$ 
& $1.3_{-0.5}^{+0.2}$ 
& - 
& $0.99_{-0.06}^{+0.06}$ 
& $0.83_{-0.08}^{+0.58}$ 
& $1.29_{-0.26}^{+0.17}$ \T\B \\

$T_\mathrm{e,s}\;\mathrm{[keV]}$ 
& $61_{-9}^{+9}$ 
& $86_{-23}^{+11}$ 
& $60_{-20}^{+20}$ 
& - 
& $47_{-8}^{+9}$ 
& $75_{-16}^{+16}$ 
& $76_{-25}^{+18}$ \T\B \\

\rowcolor{gray!15}$\tau_\mathrm{s}\;[-]$ 
& $3.29_{-0.18}^{+0.12}$ 
& $2.8_{-0.6}^{+0.5}$ 
& $1.9_{-0.7}^{+0.8}$ 
& - 
& $3.25_{-0.30}^{+0.16}$ 
& $3.22_{-0.27}^{+0.19}$ 
& $2.6_{-0.8}^{+0.6}$ \T\B \\

$T_\mathrm{in}\;\mathrm{[keV]}$ 
& $0.352_{-0.003}^{+0.004}$ 
& - 
& - 
& - 
& $0.413_{-0.007}^{+0.008}$ 
& $0.336_{-0.003}^{+0.004}$ 
& - \T\B \\

\rowcolor{gray!15}$R_\mathrm{in}\;\mathrm{[km]}$ 
& $32.6_{-0.5}^{+0.3}$ 
& - 
& - 
& - 
& $19.8_{-0.7}^{+0.8}$ 
& $32.5_{-0.7}^{+0.4}$ 
& - \T\B \\

$\alpha_\mathrm{DU1}\;[-]$ 
& - 
& $1.10_{-0.08}^{+0.05}$ 
& $1.04_{-0.12}^{+0.10}$ 
& - 
& - 
& - 
& $0.99_{-0.08}^{+0.11}$ \T\B \\

\rowcolor{gray!15}$\alpha_\mathrm{DU2}\;[-]$ 
& - 
& $1.13_{-0.08}^{+0.05}$ 
& $1.02_{-0.12}^{+0.11}$ 
& - 
& - 
& - 
& $1.01_{-0.09}^{+0.11}$ \T\B \\

$\alpha_\mathrm{DU3}\;[-]$ 
& - 
& $1.10_{-0.08}^{+0.05}$ 
& $1.02_{-0.12}^{+0.11}$ 
& - 
& - 
& - 
& $0.99_{-0.08}^{+0.11}$ \T\B \\

\rowcolor{gray!15}$\chi_0\;\mathrm{[rad]}$ 
& - 
& $-0.44_{-0.21}^{+0.18}$ 
& $-0.46_{-0.09}^{+0.10}$ 
& - 
& - 
& - 
& $-0.6_{-0.5}^{+0.6}$ \T\B \\

\end{tabular}
\end{table*}

\subsection{NICER-only analysis}\label{sec:results_nicer}
For this analysis 2000 live points and an \ac{X-PSI} sampling efficiency of 0.1 were used. Another run of 5000 live points was attempted and intermediate results were converging to the same parameters as were found here. We deem the current settings are thus sufficient and that the 2000-live-point run has converged. The 5000-live-point run was not completed due to expected high computational cost.

\begin{figure*}
    \centering
    \includegraphics[
        width=\linewidth,
        trim=3cm 0 3cm 0,
    ]{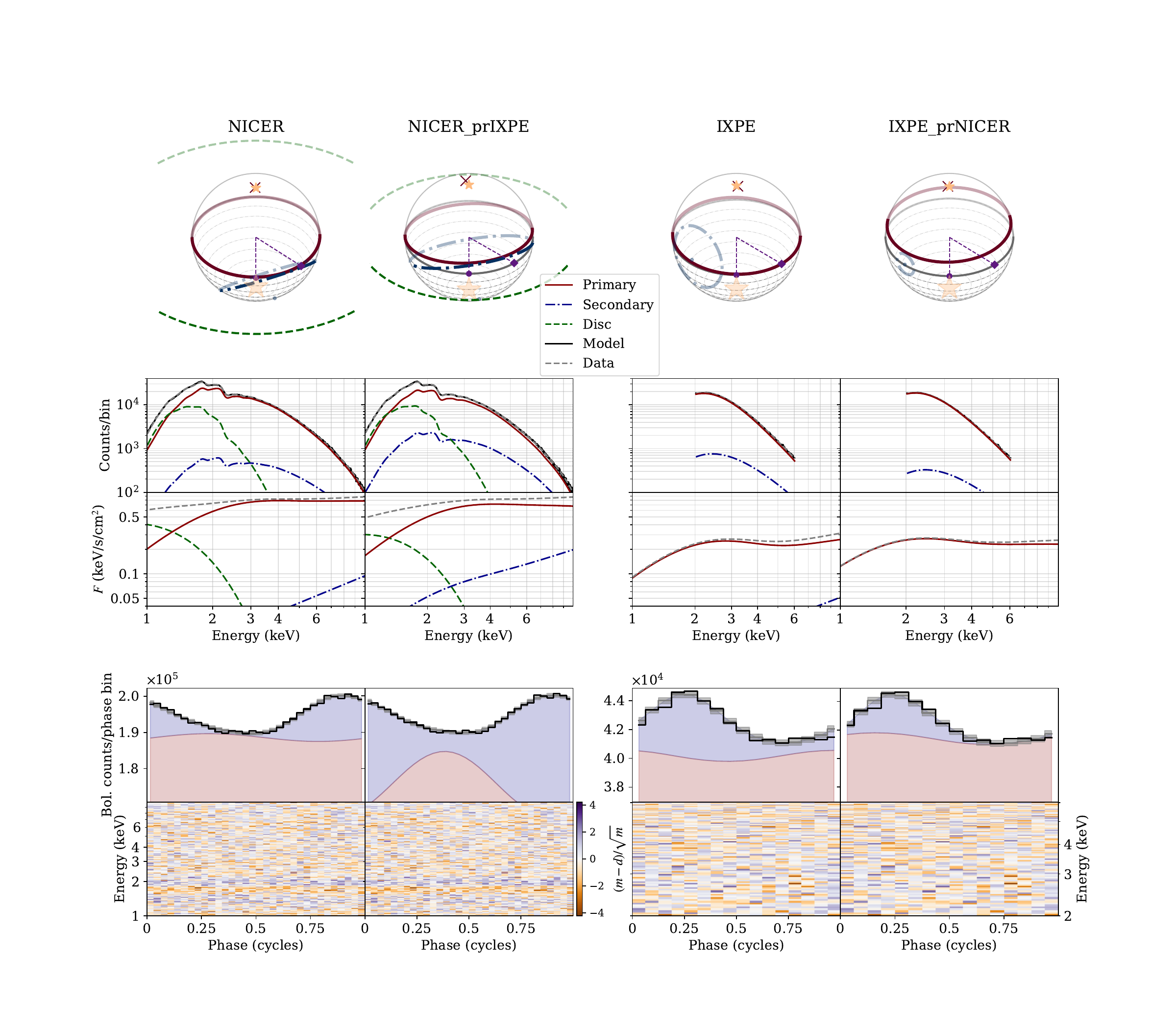}
    \caption{Projection plots, spectra, pulse profiles and residuals for the (from left to right) \ac{NICER}-only, \ac{IXPE}-informed \ac{NICER}, \ac{IXPE}-only, \ac{NICER}-informed \ac{IXPE} analyses. The \textbf{top row of panels} show the projection plots of the \acl{MAP} (MAP) samples from the perspective of the observer. The red solid lines indicate the primary hotspots, the blue dash-dotted lines indicate the secondary hotspots, and the green dashed lines indicate inner disc radii. The crosses/dots indicate the centres of the hotspots, which are on the observer/opposite side of the star, respectively. The accretion disc is not present in the \ac{IXPE} models. Relativistic effects are not shown in these projection plots. Stars indicate the northern and southern rotational poles, and the magenta circle and purple diamond indicate the phase coordinates 0 and $\pi$/4 rad, respectively. The \textbf{second row of panels} show the observed phase-summed spectra of the \ac{MAP} samples. Here, the black lines indicate data spectra, and the grey dashed lines indicate the posterior-expected model spectra. The other lines are the same as in the top panels. The \textbf{third row of panels} show the intrinsic source spectra before circumstellar and interstellar absorption and before convolving with the instrument response. The lines correspond to the same component as the above panels. The \textbf{fourth row of panels} show the bolometric pulse profile of the data and posterior-expected model in black and gray, respectively. The model also shows the 16 and 84~per cent Poisson error \acp{CI} in light gray. The blue filled-in region displays the bolometric contribution of the secondary hotspot, and red region the upper part of the contribution of the primary hot spot (and disc, if present). Finally, the \textbf{bottom row of panels} show the phase-energy resolved residuals, where $m$ is the model counts and $d$ is data counts in each phase-energy bin. In the case of the \ac{IXPE} analyses, only the spectra, pulse profiles, and residuals of DU$_1$ are shown (because DU$_2$ and DU$_3$ look very similar), although all three were fitted simultaneously during sampling.}
    \label{fig:diagnostics}
\end{figure*}

As shown in \Cref{fig:diagnostics}, analysis of the \ac{NICER} data leads to a visually good fit, apart from spectral residuals at $\sim$1.7 keV that are likely due to unmodelled \ac{NICER} spectral edges: Aluminium K edge/fluorescence (from the detector window) at 1.56 keV and Silicon K edge (from the window and bulk detector) at 1.84 keV.\footnote{\url{https://heasarc.gsfc.nasa.gov/docs/nicer/analysis_threads/arf-rmf/}} Because of these \ac{NICER} spectral edges, formally the fit is very poor, with $\chi^2$/DOF = 30108/28781 = 1.0461, corresponding to a p-value of 2.4792$\times10^{-8}$. The p-value quantifies the probability of obtaining a data set with a $\chi^2$ value at least this large, given the model with these parameters ($\chi^2=\sum_{i}(m_i-d_i)^2/m_i$, where $d_i$ and $m_i$ are data and model counts in each phase-energy resolved bin $i$ - more specifically, $m_i$ are the \ac{posterior}-expected model counts based on 100 draws from the \ac{posterior}).

Three of the parameters ($M = 2.17_{-0.03}^{+0.02}$M$_\odot$, $R_{\rm eq}=13.82_{-0.14}^{+0.10}$~km and $\zeta_{\rm p}=88.6_{-1.2}^{+0.9}\degr$) tend towards the maximum of the prior, whereas others ($D=2.88_{-0.04}^{+0.05}$~kpc, $i=50.9_{-0.5}^{+0.8}\degr$, and $\theta_{\rm p}=1.1_{-0.2}^{+0.3}\degr$) tend towards the minimum of the prior (see Fig.~\ref{fig:cornerplot_kde_combination}). It is notable that many parameters peg to the bounds of their respective \acp{prior}, which could be a hint that parameter space outside the current prior distributions are worth exploring. We discuss this possibility in \Cref{sec:discussionpriors}. The corresponding spot geometry has a large, bright primary hotspot that produces a significant number of non-pulsed counts, while wobbling slightly around the rotational axis, as shown by the projection in \Cref{fig:diagnostics}. This geometry is very similar to what was found in \citet{Dorsman2025} for SAX~J1808.4$-$3658. \ac{rin} is also very large, $32.6^{+0.3}_{-0.5}$~km and near $R_{\rm co}$, the upper bound of the prior. In combination with the inferred \ac{i}, this implies that no rays from the secondary hotspot are blocked from the observer. 

The spectrum in \Cref{fig:diagnostics} shows that the accretion disc dominates at low energy but becomes subdominant at 1.3 keV. Above $\sim$2 keV, the primary hotspot is responsible for over 90~per cent of the counts. The secondary hotspot, despite the fact that it creates considerably fewer counts than the primary, is almost entirely responsible for shape of the pulsation.

\subsection{IXPE-only analysis}\label{sec:results_ixpe}
The \ac{IXPE} data consisted of Stokes $I$, $Q$, and $U$ data. For comparison, we performed \ac{IXPE}-only analyses both with and without fitting the Stokes $I$ data. Unlike the \ac{NICER} analysis, for the final \ac{IXPE} analyses the accretion disc model was not applied. We did a test including the accretion disc for the Stokes $IQU$-analysis, but the run was much slower and resulted in a worse evidence than a run without the disc with the same sampler settings (1000 live points and 0.3 sampling efficiency). Presumably good fits are found without the disc because its contribution to the \ac{IXPE} counts is negligible given \ac{IXPE}’s harder energy range compared to \ac{NICER}.

For the $IQU$ results presented in Table \ref{tab:inferences} and Fig. \ref{fig:cornerplot_kde_combination}, we used 4000 live points and 0.1 sampling efficiency. 
With the initial sampler settings, the mode corresponding to a smaller \ac{r} and and higher \ac{m} was not recovered (see Fig. \ref{fig:cornerplot_kde_combination}). 
The mode appeared when increasing the number of live points from 1000 to 4000, whereas reducing the sampling efficiency from 0.3 to 0.1 did not lead to significant further shifts in the posteriors. 
Nevertheless, slight broadening of many credible intervals was still observed, and the convergence was thus not formally proven. 
That would have been computationally expensive to do, because the last run spent already almost 50 kilo core-hours (\acsu{kch}). 
For the Stokes $Q$ and $U$-only analysis, further improved sampler settings were used (10000 live points and 0.1 sampling efficiency), due to the very short run time (less than 1 \ac{kch}) of the much noisier data. 
The full posterior distributions from these runs are shown in Appendix \ref{sec:appendix_extra_corner_plot}. 

The $QU$ and $IQU$ analyses result in visually good fits in all the fitted phase-resolved Stokes spectra (see e.g. Stokes $I$ for DU$_{1}$ in \Cref{fig:diagnostics}). For the Stokes $I$ fit statistics (combined over detector units) we get $\chi^2$/DOF = 4940/4827, corresponding to a p-value of 0.12632. As in \citet{Papitto2025}, the $IQU$ fit is dominated by the Stokes $I$ signal since the polarization degree is rather low ($< 5 \%$) and the polarization angle is roughly constant in phase.

The results generally agree with those obtained in \citet{Papitto2025}. In particular, the inferred geometry matches that of \citet{Papitto2025} when we analyse only the $Q$ and $U$ data, although with large credible intervals. In this case, we find, for example, that $i=77^{+8}_{-10}\degr$, $\theta_{\mathrm{p}}=17^{+50}_{-9}\degr$, and $\theta_{\mathrm{s}}=156^{+17}_{-50}\degr$. Slightly different constraints are obtained when analysing all $I$, $Q$, and $U$ data, with bimodal posterior distributions in many parameters.\footnote{Preliminary tests fitting only Stokes $I$ yielded results very similar to those obtained when fitting all Stokes parameters.} In this case, we find $i=63^{+12}_{-8}\degr$, $\theta_{\mathrm{p}}=0.7^{+0.4}_{-0.3}\degr$, and $\theta_{\mathrm{s}}=146^{+13}_{-10}\degr$, and notably a larger primary hotspot size, with $\zeta_\mathrm{p}$ covering almost half of the NS surface (similar to the \ac{NICER} analysis). In both cases the NS mass and radius are only weakly constrained. 

\subsection{Combination method 1: combining independent posteriors}\label{sec:results_kde}
This section presents the results of combining the \acp{posterior} of the \ac{IXPE}-only (Stokes $I$, $Q$, and $U$) with the \ac{NICER}-only analysis. We used \acp{KDE} to add the ln-likelihood surfaces, following the methodology outlined in \Cref{sec:inference}. As a reminder, here only the constant parameters (\ac{tc}) are combined, because all potentially variable parameters may change between the recording of the two data sets. For the final sampling step, we used a sampling efficiency of 0.1 and 4000 live points.

\begin{figure}
    \centering
    \includegraphics[width=\linewidth]{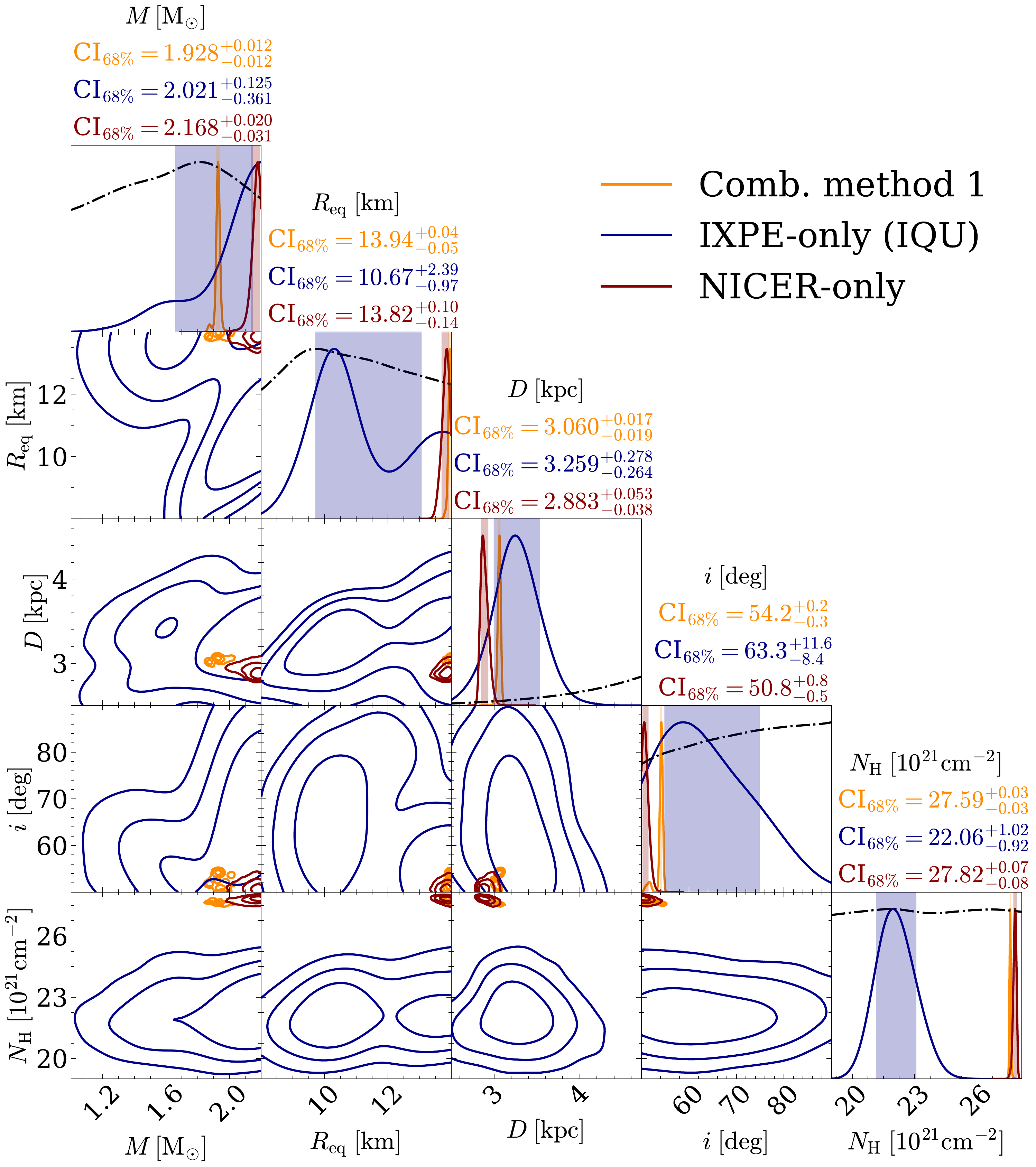}
    \caption{Posterior distributions of the constant parameters from the \ac{NICER}-only and \ac{IXPE}-only analysis, and their \ac{KDE} combination (method 1, see \Cref{sec:results_kde}). The dashed and solid lines on the diagonal show the \ac{1D} marginal \acp{prior} and \acp{posterior} of each parameter, respectively. For visibility, these are scaled such that the distributions peak at the same height. The coloured intervals correspond to the 68.3~per cent \acp{CI} also listed above the diagonal. The off-diagonal plots show 2D marginal \acp{posterior} for each parameter combination. There, contours correspond to the 68.3, 95.4, and 99.7~per cent credible regions.}
    \label{fig:cornerplot_kde_combination}
\end{figure}

\Cref{fig:cornerplot_kde_combination} shows a corner plot of the inferred \acp{posterior} of \ac{tc}. Interestingly, the combined run results in a lower $M=1.928_{-0.012}^{+0.012}$ M$_\odot$ and higher \ac{r}$=13.94_{-0.05}^{+0.04}$ km than either of the individual runs. At first glance this result seems counter-intuitive, but we interpret it by looking at the \ac{m}$-$\ac{r} \acp{posterior}. Here, we see that the \ac{NICER}-only \ac{posterior} lies in the top-right corner, whereas the \ac{IXPE}-only posterior disfavours this corner in favour of either lower \ac{m} or lower \ac{r}. It appears that the best overlap found during sampling was towards lower \ac{m} and thus the combined posterior is found there. The \acp{CI} of the combination are noticeably narrower than \acp{posterior} of either constituent runs, as is expected from the multiplication of two peaked distributions, and more so if they are partially disjoint.

Based on the change in \ac{nh}, we consider the possibility that \ac{nh} is not a constant parameter that, instead, has decreased between the \ac{NICER} to the \ac{IXPE} observations. To cover for that possibility, we exclude the \ac{nh} from the combination and rerun the final sampling step in the combination process. This moves the \acp{posterior} significantly (beyond their 68~per cent \acp{CI}), with now $M=1.885_{-0.015}^{+0.020} M_\odot$ and \ac{r}$=13.84_{-0.06}^{+0.05}$ km, although this change is small compared to the width of both the \ac{NICER} and \ac{IXPE}-only \acp{posterior}.  

\subsection{Combination method 2: two-stage Bayesian analysis}\label{sec:results_posterior_informed}
This section describes the two-stage Bayesian analyses, where the posteriors of the constant parameters that result from the analysis of one data set are used as \acp{prior} in the analysis of the other data set. This analysis is done in two ways: using \ac{IXPE}-only ($IQU$) \acp{posterior} as \acp{prior} for subsequent \ac{NICER} analysis, and vice versa.

\subsubsection{IXPE-informed NICER analysis}
In this analysis the \acp{posterior} that resulted from the \ac{IXPE}-only analysis were implemented as updates to the \acp{prior} of the constant parameters for the \ac{NICER} analysis. This analysis was done with 2000 live points and a sampling efficiency of 0.1, like the \ac{NICER}-only analysis. Compared to 1000 live point runs, these results did not change much, indicating that the sampler settings were sufficient for these runs to be converged. 

As shown in \Cref{tab:inferences}, \ac{m} is consistent with the \ac{NICER}-only result at $2.15^{+0.03}_{-0.06}$ M$_\odot$, but \ac{r} shrinks significantly to $11.1^{+0.3}_{-0.3}$ km and is consistent with the \ac{IXPE}-only result. \ac{rin} shrinks to $19.8^{+0.8}_{-0.7}$km, which implies a brighter disc and the partial occlusion of the emission of the secondary hotspot. The second column in \Cref{fig:diagnostics} shows that the hotspot configuration closely resembles the configuration found in the \ac{NICER}-only analysis. The secondary hotspot is brighter compared to \ac{NICER}-only, owing to its higher \ac{tbb} and \ac{zeta}.  Unlike before, we can now clearly see that the reduction in flux between NICER and IXPE observations (see Fig. \ref{fig:diagnostics}) is mainly due to the decrease in $T_{\rm seed,p}$. This is consistent with the idea that the accretion rate is higher during the \ac{NICER} observation. 
 
$\mathcal{Z}$ is significantly worse for this model, with a penalty of 41.19 in ln-space \citep[see e.g.][]{kass1995bayes}, and the fit is also significantly worse, with the $\chi^2/$DOF increasing to $30204/28781 = 1.0494$ and the p-value reducing by around a factor of 10 to 2.6326$\times 10^{-9}$. These metrics indicate that this solution is significantly disfavoured by the \ac{NICER} data compared to the \ac{NICER}-only solution.

To allow for the possibility that \ac{nh} is not a constant parameter (but evolves during the outburst), we also ran the \ac{IXPE}-informed \ac{NICER} analysis while keeping the \ac{nh} prior from \Cref{tab:parameters}. The solution found in this case is better than the previous \ac{IXPE}-informed \ac{NICER} result, and marginally better than even the \ac{NICER}-only result, as shown by $\ln(\mathcal{Z})$ and the $\chi^2$ value in \Cref{tab:inferences}. This finding supports the idea that \ac{nh} might not be constant, but instead may have decreased between the time of the \ac{NICER} observations and the \ac{IXPE} observations. Compared to the previous \ac{IXPE}-informed \ac{NICER} result, \ac{m} is effectively unchanged, but \ac{r} increases to $13.6_{-0.3}^{+0.3}$ km and is now more consistent with the \ac{NICER}-only result. This improved consistency is also true for the other parameters, such as \ac{nh} and the hotspot configuration. This experiment shows that a good prior on \ac{nh} is important for the inference of \ac{r}. 

\begin{figure}
    \centering
    \includegraphics[width=\linewidth]{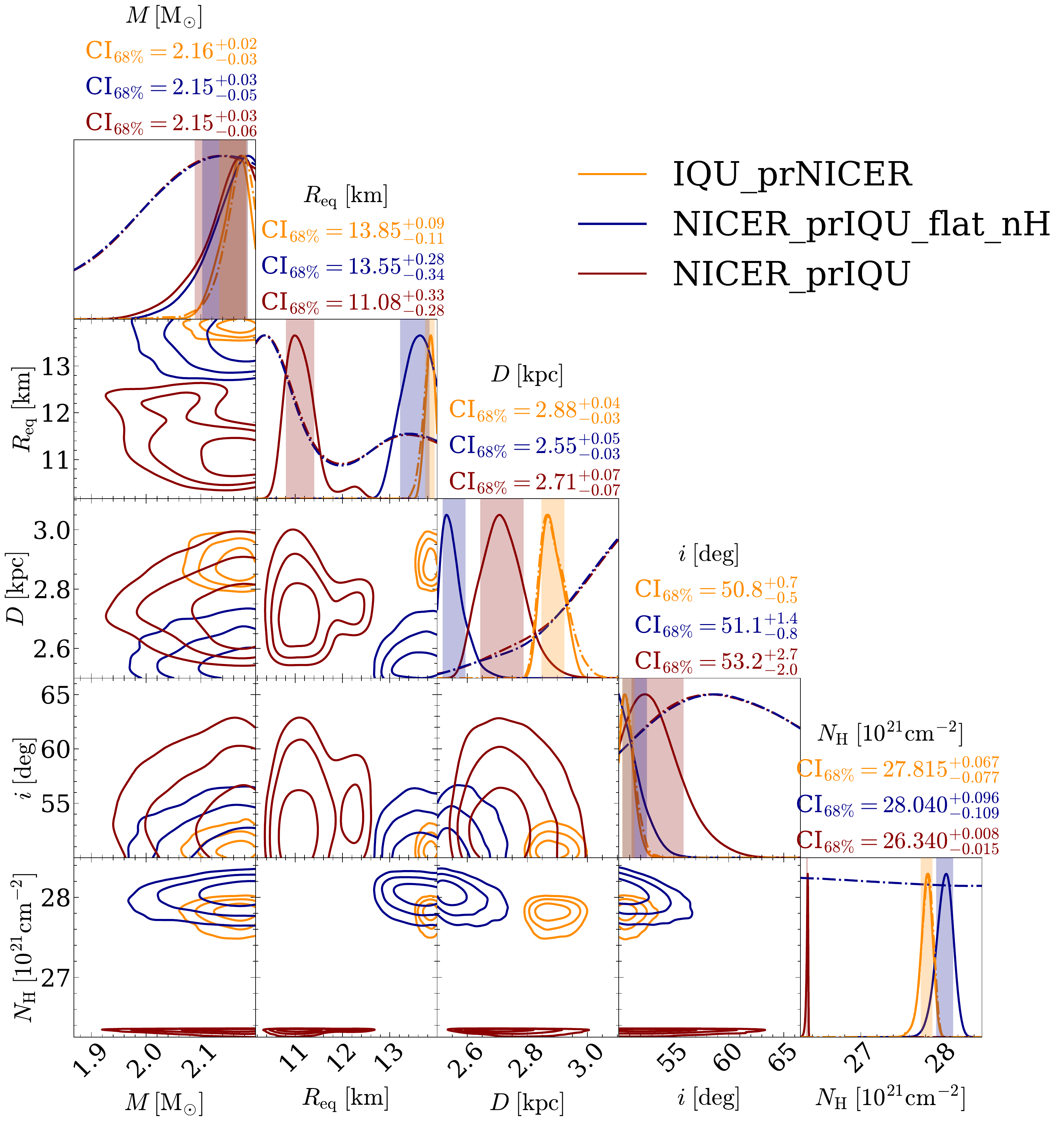}
    \caption{Posterior distributions for the \ac{IXPE} Stokes $I$, $Q$, and $U$ analysis informed by \ac{NICER} \acp{posterior} (IQU$\_$prNICER), and two \ac{NICER} analyses informed by \ac{IXPE} \acp{posterior}. The first retains the flat \ac{nh} prior (NICER$\_$prIQU$\_$flat$\_$nH) and the second utilises the \ac{IXPE} posterior of \ac{nh} as prior (NICER$\_$prIQU). These correspond to combination method 2 (See \Cref{sec:results_posterior_informed}). See the caption of \Cref{fig:cornerplot_kde_combination} for a detailed description of the contents of the diagonal and off-diagonal panels.}
    \label{fig:cornerplot_posterior_informed_priors}
\end{figure}

\subsubsection{NICER-informed IXPE analysis}\label{sec:results_ixpe_nicer-informed}
In this analysis the posteriors that resulted from the \ac{NICER} analysis were implemented as updates to the priors of the constant parameters for the \ac{IXPE} ($IQU$) analysis. While the number of live points is still 4000, we note though that for computational speed-up, the sampling efficiency was set in this case to 0.3 instead of 0.1.

Interestingly, the $\ln( \mathcal{Z})$ found here is 3 higher in ln-space than in the \ac{IXPE}-only analysis, which is notable but not definitive evidence in favour of this model. However, the $\chi^2$/DOF fit is marginally worse at 4953/4827 = 1.0262. As shown in \Cref{tab:inferences}, the constant parameters are mostly consistent with the updated \ac{prior} (the \ac{NICER}-only \ac{posterior}). Here, \ac{m} is inferred to be $2.16^{+0.02}_{-0.03}{\rm M}_{\odot}$ and \ac{r} is inferred to be $13.85^{+0.09}_{-0.11}$~km. Similarly as for the \ac{IXPE}-informed \ac{NICER} analysis, we see that the reduction in flux compared to the \ac{NICER} observation is mainly caused by a reduction in $T_{\rm seed,p}$.

The variable parameters are mostly consistent with the \ac{IXPE}-only \acp{posterior}. \Cref{fig:cornerplot_posterior_informed_priors} shows that, as we had speculated in \Cref{sec:inference}, rather than leading to consistent results, there are differences between the results of the two-step analyses, depending on which data set is analysed first.

\section{Discussion}\label{sec:discussion}
In this section, we will discuss the obtained results and this discussion is structured as follows. \Cref{sec:discussionsystemparams} starts out by providing an overview of obtained system parameters and compares them to the literature where possible. Multiple methods of data analyses were explored -- in which the two data sets were considered both separately and together -- so, \Cref{sec:reliability} discusses reliability and fit quality of each analysis. Finally, \Cref{sec:recommendations} gives recommendations for future work.

\subsection{Overview of inferred parameters and literature comparison}\label{sec:discussionsystemparams}
This section provides an overview of inferred parameters. These are grouped into parameters that govern hotspots, mass and radius, geometry, pulse profile, distance, the polarization and the spectrum. We compare results against literature values where possible.  

\textbf{Hotspots:} a consistent trend in the results is that the hotspots of \SRGA{} are very large: the primary hotspot occupying up to half of the stellar surface, and the secondary hotspot size in the range $\sim$20$-$70\degr. We can compare this against \citet{Kajava2011} and \citet{Salmi2018}, who found smaller hotspots for SAX~J1808.4$-$3658. They used \textit{RXTE} data that covers a higher energy range, however, where the effective hotspots could look smaller if they have a temperature gradient. In addition, they accounted for disc reflection in their models, which we do not, and that might reduce their inferred hotspot size. \citet{Dorsman2026} also find generally large hotspots for SAX J1808.4$-$3658 using \ac{NICER} data, although they were not as large when they used the \textsf{STU-Disc} configuration: $\zeta_{\rm p} \approx$ 47\degr--59\degr. Simulations \citep[e.g.][]{Romanova2004,Kulkarni2013,Das2025} predict typically smaller hotspots, but perhaps more importantly they predict very different hotspot shapes than those allowed in our model space, e.g. crescents, rings, and elongated shapes. Although upgrades to hotspot shape modelling are needed, our results indicate that the `effective' hotspots of \acp{AMP} could cover a significant portion (even up to half) of the stellar surface. 

\textbf{Mass and radius:} the analyses presented here consistently yielded a high \ac{m} at 1.9--2.2\ M$_\odot$, where 2.2 M$_\odot$ is the upper bound of the prior. For \ac{r}, we observe that generally two solution modes are found. On the one hand there is a high \ac{r} mode (13.4--14~km), associated with the \ac{NICER}-only analysis, analyses with the \ac{NICER} prior and the \ac{KDE} combination. On the other hand there is a low \ac{r} mode (9.7--13.1~km), preferred by the \ac{IXPE}-only analysis and analysis with the \ac{IXPE} prior (although \Cref{fig:cornerplot_kde_combination} shows that the \ac{IXPE}-only analysis also found a high radius mode). The inferred \ac{m} and \ac{r} are near the prior boundaries in many cases, so we recommend exploration of the robustness of results against changes in \ac{m}-\ac{r} priors. We discuss this point further in \Cref{sec:discussionpriors}. We recommend this additional step should be done before including these present results in any new \ac{EoS} inference.

\textbf{Geometry:} taking our results together, we find the inclination of the system to be moderately high ($i=$50\degr--75\degr), the primary hotspot to be centred closely around the rotational axis ($\theta_p=$0\degr--4\degr), and the secondary centred on the opposing hemisphere ($\theta_{\rm s}=$125\degr--159\degr). We note here that $i$ at 50\degr is close to the lower boundary prior boundary, which was only based on inferred values in literature, so this motivates an investigation with a lower prior in future work. Furthermore, we found \ac{rin} to be distributed in the range of 19--33~km.

Let us compare these results against \citet{Molkov2024discovery}, who investigated the persistent pulsations discovered in the \ac{ART-XC} data, which started just before and overlapped with the \ac{NICER} observation. \citet{Molkov2024discovery} fitted the pulse profile by eye, assuming two antipodal spots and fixed \ac{m} $=1.4\;{\rm M}_{\odot}$ and \ac{r} $=12$ km. They used a simplified atmosphere model $I(\alpha) \propto 1 + h \cos \alpha$ where $\alpha$ is the angle relative to the local normal in the spot co-moving frame and $h$ is the anisotropy parameter $-1<h<0$. They found a good fit with \ac{i} $=58\degr$, \ac{theta} $=14\degr$, \ac{zeta} $=33\degr$, and \ac{rin} $=24.6$ km. While their \ac{i} and \ac{rin} are in agreement with our results, their \ac{theta} and \ac{zeta} are not. This disagreement may be accounted for by their restriction to antipodal spots, simplified atmosphere model, and limited search of parameter space.

Let us also compare the results against \citet{Papitto2025}, who performed simplified pulse profile modelling on the \ac{IXPE} Stokes $I$, $Q$, and $U$ data. They assumed two anti-phased point-like spots with an emission model consisting of power-law emission $I(E) \propto E^{-(\Gamma-1)}$, $\Gamma$ fixed at 1.7, and same simplified anisotropy distribution with parameter $h$ as \citet{Molkov2024discovery}. They also used a normalisation factor to scale the flux to the data. Furthermore, they restricted their phase-resolved analysis to the 3--6~keV energy band, combining all energy channels within this interval to maximize the polarized signal, whereas we performed a fully phase- and energy-dependent analysis over the 2--6~keV range. Using fixed $M=1.4\,{\rm M}_\odot$, $R=10$ km, and $h=-0.3$, they found $i = 74.1^{+5.8}_{-6.3}\degr$, which is consistent with our \ac{IXPE}-only Stokes $I$, $Q$ and $U$ result, but inconsistent with any of the results where \ac{NICER} data were involved (in which $i$ is closer to 50\degr). Furthermore, they infer $\theta_p = 11.8^{+2.5}_{-3.5}\degr$, and $\theta_s = 172.6^{+2.0}_{-1.0}\degr$, which are somewhat in tension with our results, except the \ac{IXPE} $Q$ and $U$ analysis (although all our inferred hotspots lie in the same general region on the NS surface). These differences in results can easily be accounted for by the various differences in modelling mentioned above. 

\textbf{Pulse profile and timing properties:} in \citet{Molkov2024discovery}, the primary hotspot causes the main pulse and the secondary causes the inter-pulse, with a plateau due to partial eclipse by the accretion disc. \ac{NICER} also observed the plateau (see \Cref{fig:pulseprofiles}), so it is unlikely that this feature can be attributed to Poisson noise. Motivated by \citet{Molkov2024discovery} and unlike \citet{Dorsman2026}, we have included occlusion of hotspot emission by the disc. However, our best fit to \ac{NICER} data does not require disc blocking and still leaves some curvature in the pulse profile where the plateau should be. We judge our fit to be acceptable, because we see no clear phase-dependent structure in the residuals, despite that difference. Still, the \ac{IXPE}-informed \ac{NICER} fit (second column in \Cref{fig:diagnostics}) features a partially occulted secondary hotspot. In contrast, here the primary is in anti-phase with the overall pulsation and is responsible for flattening the inter-pulse. In the end, we judge that the results are not conclusive and more investigation is warranted. \citet{Li2025Timing} also studied the pulsations and revised the spin evolution, favouring a spin-up episode during the \ac{NICER} observation. They selected a longer time interval for this timing solution however, that we exclude due to high undershoot rates in the \ac{NICER} data (see Section \ref{sec:dataprep_nicer}).

\textbf{Distance:} we found the system to be very close-by (2.5--3.3~kpc) compared to what has so far been inferred in the literature ($\sim$8--11~kpc, see \Cref{sec:priors}). We advise caution when using this value range, since we infer $D$ near the lower prior bound. The lower bound was selected rather arbitrarily to cover the lower end of the distance estimate range of \citet{Molkov2024discovery}, so we do not associate our lower bound with a high level of confidence. We recommend the adoption of a more physically-motivated lower prior bound in the future, if available. A smaller distance corresponds to a brighter signal, so we speculate that the discrepancy with the literature is explained by us omitting a source of radiation in the model, such as the accretion column or disc reflection.

\textbf{Absorption column:} the analyses of the \ac{NICER} and \ac{IXPE} data inferred different values for \ac{nh}: $(22.1^{+1.0}_{-0.9}) \times 10^{21}$ cm$^{-2}$ for \ac{IXPE}-only or (26--28)$\times 10^{21}$ cm$^{-2}$ if \ac{NICER}-data are in any way included in the analysis. This may indicate that this parameter has evolved between these observations. This variation is an important consideration, because in the combined analyses, we found that treating this parameter either as constant or variable influences the inference of \ac{r}.

\textbf{Polarization:} \citet{Papitto2025} suggest that the most natural explanation for the polarization detected by \ac{IXPE} is Comptonised emission from hotspots, with \ac{te} $\sim$20$-$60~keV, \ac{tbb} $\sim$1~keV, and moderate optical depth at \ac{tau} $\approx$1--2. This origin is mostly consistent with our findings in the analysis of \ac{IXPE} data. We found \ac{te} ranging from 37 to 97 keV, \ac{tbb} ranging from 0.65 to 1.5 keV, and \ac{tau} from 1.1 to 1.4. The primary hotspot is typically towards the lower end of the range for these three parameters while the secondary is towards the higher end.

\textbf{Spectrum:} the broadband X-ray spectrum was studied by \citet{Malacaria2025Disk} with overlapping \textit{XMM-Newton} and \textit{NuSTAR} data. They found that an absorbed \texttt{bbodyrad} ($T_{\rm bb}= 0.64^{+0.02}_{-0.01}$ keV) and \texttt{diskbb} ($T_{\rm in}=0.44^{+0.09}_{-0.06}$ keV) and \texttt{nthComp} ($T_{\rm seed}=0.93^{+0.02}_{-0.02}$ keV) account well for the lower energy part ($\lesssim$ 4 keV) of the spectrum. These observations took place just after the \ac{NICER} observation, at the start of the \ac{IXPE} observation. Our \ac{IXPE} pulse profile is averaged over a longer part of the decline, so the results are not fully comparable. Still, we find some agreement. Their \texttt{diskbb} $T_{\rm in}$ is only a bit higher than our \ac{tin} ($0.3-0.4$ keV). Their \texttt{nthComp} $T_{\rm seed}$ is similar to our $T_{\rm bb,p}$ and $T_{\rm bb,s}$ during the \ac{NICER} observation, although $T_{\rm bb,s}$ increases in the later \ac{IXPE} observation. Their $T_{\rm e}$ is a lower temperature ($18.2^{+0.7}_{-2.7}$ keV), which is close to our findings for $T_{\rm e,p}$, but not for $T_{\rm e,s}$, where our inferred value is significantly hotter. The differences could perhaps be explained by their more detailed modelling of reflection, which starts to become dominant at higher energy but is not accounted for in our model. Also notable is that they have another low energy \texttt{bbodyrad} component, which we do not have. Plausibly such a component could arise if the surface of the \ac{NS} outside of the hotspots is hot enough. 

For the higher energy ($\gtrsim 4$ keV) part of the spectrum, they found good fits by adding to the aforementioned lower energy components either a semi-phenomenological model that features two \texttt{diskline} models or a comprehensive reflection model that accounts for the broadened iron line and other reflection from the disc. They infer different values between these models for $i$ and $R_{\rm in}$. Their model with two \texttt{diskline} components fits best at $i=85\degr$ and $R_{\rm in}= 10.0^{+1.2}_{-1.7} R_{\rm g}\,(30_{-4}^{+5}\,M_{2}$ km, where $M_{2} = 2\,{\rm M}_\odot$). However, their reflection model fits best with $i=53\degr$ and $R_{\rm in}=6 R_{\rm g}\,(18\,M_{2}$ km), the latter being pegged at the lower limit. Our results are more in agreement with the smaller result for \ac{i}, but both inferred values for \ac{rin} are consistent with our results, since we found both low and high values.

\citet{Li2025Timing} also analysed the X-ray broadband spectrum using \ac{NICER}, \textit{Einstein Probe}, \ac{IXPE}, \textit{Insight-HXMT}, and \textit{INTEGRAL} data. They found good spectral fits using absorbed \texttt{nthcomp}+\texttt{gaussian} in their spectral model to account for the \ac{NICER} systematic at 1.7 keV. However, their \ac{tbb} $\sim$ 0.4 keV and \ac{te} $\sim$ 4--6~keV are much colder than our \ac{tbb} $\sim$ 0.8--1.1~keV and \ac{te} $\sim$ 22--89~keV inferred for the \ac{NICER} data. These diverging results are likely due to model choice, because we separately include a contribution from the disc blackbody at lower temperature. In their best fit model \texttt{nthcomp}+\texttt{gaussian}+\texttt{relxillcp}, where the latter accounts for the broad band accretion disc continuum and reflection, they still find similar \ac{tbb} and \ac{te} as before.

\subsection{Reliability and fit quality}\label{sec:reliability}
Here, we discuss the reliability and, if applicable, fit quality of the results of each analysis. This discussion is subdivided as follows:  \Cref{sec:discussionNICERonly} discusses the results of the stand-alone analysis of the \ac{NICER} data, and \Cref{sec:discussionIXPEonly} the same of the \ac{IXPE}-data. \Cref{sec:discussion_joint} discusses the results of the combined data analyses.

\subsubsection{PPM with NICER only}\label{sec:discussionNICERonly}
In this analysis, no clear pulsed structure remains in the residuals, which shows that the two circular hotspot model is sufficient to fit the data. However, the overall fit-quality is not good, due to a non-pulsed structure remaining at $\sim$1.7~keV, which we attribute to unmodelled \ac{NICER} spectral edges. 

We also noted that these data lead to tight parameter constraints for the pulse profile model. However, several parameters are concentrated towards the bounds of their priors: \ac{m}, \ac{r}, $\zeta_{\rm p}$ near their respective upper bounds and \ac{d}, \ac{i}, \ac{rin} near their lower bounds. The apparent tightness of these posterior distributions may be driven by prior bounds rather than the data. Accordingly, confidence in these results depends on the validity of the priors. We recommend reanalysis using alternative models and more conservative or better-justified priors to assess robustness.

\subsubsection{PPM with IXPE only}\label{sec:discussionIXPEonly}
In this case, good fits of the Stokes $I$ data were found, indicating that the model is sufficient to fit the data, even without including an accretion disc. The parameter constraints are broader and generally overlapping with the \ac{NICER}-only constraints, except for the posterior peaks of \ac{nh} and \ac{r} which are shifted to smaller values. 

The polarization data were not constraining enough to affect the results much, compared to the constraints imposed by the \ac{IXPE} Stokes $I$ data. This finding is similar to that of \citet{Papitto2025}, where most parameter information was also set by the total pulse profile alone. This may be explained by the relatively small polarization degree and very small phase-resolved variability of the polarization angle. Possibly, this will change for other stars and for future instruments like \ac{eXTP}. The \ac{IXPE} constraints on the geometry are also broader compared to the \ac{NICER}-only constraints.

When comparing the analyses of the \ac{IXPE} data versus the \ac{NICER} data, it is clear that \ac{NICER} data is more constraining. However, we consider the \ac{IXPE} constraints important despite being broader than the \ac{NICER} \acp{posterior}, because these evade potential biases specific to the analysis of \ac{NICER} data, such as those related to systematics or the accretion disc \citep[the latter was reported on by][]{Dorsman2025}. The Stokes $Q$ and $U$ results (see \Cref{fig:cornerplot_IXPEonly_pol,fig:cornerplot_IXPEonly_rest}) are a relevant point of comparison, meanwhile, if one is interested to avoid potential bias that may be caused by inaccurate modelling of the non-polarized emission.

\subsubsection{PPM with combined NICER and IXPE data}\label{sec:discussion_joint}
We have seen that the two data sets give different answers on their own, even for parameters that should be the same, such as \ac{m}. This discrepancy is intriguing and showcases that joint analysis is helpful to point out parameter evolution or inconsistency in inferred parameters. We explored two different methods to combine the data: (1) \ac{KDE}-based combination of separate analyses, and (2) two-step Bayesian analyses. For discussion of the third method: a joint model, we refer to \Cref{sec:appendix_sts}, where this approach is explored to a limited degree.

Combination method 1 takes the samples of the two separate analyses and estimates the likelihood surfaces from them using \ac{KDE}. A new sampling analysis is then done on the sum to estimate the final joint posterior. We found this method to be fast and affordable: only a few parameters, those assumed constant between observations, are used in the final sampling and the likelihood evaluation is fast. However, the results obtained with the KDE-based combination method may depend on the specific KDE hyperparameter choices used to reconstruct continuous likelihood surfaces from the two separate runs (\ac{NICER} and \ac{IXPE}), which are then combined to explore the shared parameter space. In this analysis, we adopted Scott’s rule of thumb to set the KDE bandwidth~\citep{Scott1992}. 
If we assume that \ac{nh} is a constant parameter, the differences observed between runs performed with and without tying \ac{nh} are difficult to reconcile. At least we can say that the sufficiency of kernel settings are an important consideration in this case, where the posterior distributions from \ac{NICER} and \ac{IXPE} are very disjoint. The reason is that the overlap region -- where the joint posterior gets support -- may be less accurately captured by the \ac{KDE} approximation. At the same time, the strong disagreement between the two independent \ac{nh} constraints could also reflect genuine astrophysical variability in the absorbing column density. If so, \ac{nh} should not be assumed to be constant between the two observations, and forcing a shared value would bias the combined inference.

In combination method 2 we used the \acp{posterior} of the constant parameters of a first analysis to update the priors of the second analysis. This was done both ways (\ac{IXPE} first or \ac{NICER} first) to check mutual consistency. Because the \ac{NICER} data is much more constraining than \ac{IXPE} data (due to the larger number of counts), we see that \ac{NICER}-informed \ac{IXPE} \acp{posterior} strongly resemble the priors, while this is not the case when the analysis ordering is reversed. In theory we should obtain the same results, but we find that the order matters, so this method is not robust. Possibly a higher sampling resolution would converge the two results, and whether this is true could be tested with real data or synthetic data. This would likely have a high computational cost, however. We also found that the assumption of whether \ac{nh} is a constant parameter influences the inference of \ac{r}.

\subsection{Recommendations for future work}\label{sec:recommendations}
Based on the discussion so far, we will discuss here recommendations for future work. This section is subdivided into four areas of possible improvement that we consider: the computational efficiency, model improvements (to improve physical realism), improvements to the prior, and the methodology of joint data analysis. 

\subsubsection{Computational speed-up}\label{sec:discussion_speedup}
Given the computational cost of this type of analysis, it is sensible to invest in improving the computational efficiency. Another important caveat is that, as shown by \Cref{fig:outburst}, the pulse profiles were averaged over a period of time within which the flux was changing substantially. As we have seen in \Cref{sec:appendix_sts}, accounting for variable parameters increases the computational cost substantially; this was studied in more detail by \citet{Kini2024a,Kini2024b} within the context of \acp{TBO}. 

The model evaluation of \acp{AMP} is in principle more computationally expensive compared to \acp{RMP} because of the atmosphere model, accretion disc, and polarized data. The cost here has been kept low primarily by the lack of exploration of more complex hotspot models. A straightforward way to reduce computational cost is to fine-tune the various resolution parameters and study the approximations used in \ac{X-PSI}, so as to achieve the right balance of accuracy and speed, potentially tailored for the data set at hand \citep[see also][]{AlGendy2014,Choudhury2024b,Dorsman2025,Mauviard2025}. More drastic but potentially rewarding on the longer term would be the porting of \ac{X-PSI} to \acp{GPU}, similar to the implementation by \citet{Bronzwaer2018,Zhou2025}, the usage of machine learning to compute pulse profiles \citep[e.g.][]{Chan2013,Waldrop2025} or the usage of simulation-based inference to simulate \acp{posterior} \citep[see e.g.][for a review]{Cranmer2020}.

The usage of a joint model to simultaneously fit data is an attractive strategy: it consists of only a single sampling analysis, and we attempted this in \Cref{sec:appendix_sts}. Because the observations were disjoint, we had to account for variable parameters, leading to high computational cost. As a consequence, we were limited to the usage of the \sts{} model (i.e. antipodal hotspots) to analyse the data sets, and found significantly worse fits compared to the analyses in the main body of the paper. In addition to the measures to reduce computational cost listed so far, one could alleviate the cost here by assuming variable but apparently consistent parameters to be constant. This consistency can be based on the results of other analyses, such as how the parameters $\theta_{\rm p}$ and $\zeta_{\rm p}$, were found to be in good agreement across \Cref{tab:inferences}. This will not always be an option, though, and in general we recommend restricting the analyses with joint models to (mostly) simultaneously observed data. In addition to that, the flux and pulse properties during the data selections should be preferably constant so that the reasonable assumption can be made that related parameters are constant, which may be achieved using data cuts. This reduces the computational cost by limiting the number of variable parameters, and could possibly make \stu{} computationally viable. 

\subsubsection{Improvements to the model}\label{sec:improvements}
There are many possible improvements to the model that could enhance its realism and therefore enhance the accuracy by which parameters are inferred. One major concern mentioned in \Cref{sec:reliability} was that certain emission components could be missing, and we recommend in particular improving the modelling of the accretion environment. For example, we recommend the inclusion of the extended accretion column structure, which \citet{Ahlberg2024} has shown can significantly modify the hotspot emission pattern. Furthermore, although the accretion disc model has been upgraded to block radiation from the hotspots, the disc emission model is still a basic multicolour blackbody. A step up would be a general relativistic disc model (which also includes polarized emission), such as was described by \citet{Loktev2022}. In addition, spectral modelling by e.g. \citet{Malacaria2025Disk} and \citet{Li2025Timing} has shown that disc reflection modelling is an important feature of the broadband spectrum of \SRGA, so its inclusion in \ac{PPM} is certainly warranted. Besides the modelling of the accretion environment, it was mentioned in \Cref{sec:discussionsystemparams} that the circular hotspot shapes used in this work are too simplistic compared to results from simulations. Potential improvements to the modelling of hotspot shapes for \acp{AMP} was discussed in more detail in section 6.4 of \citet{Dorsman2025}.

As also stated in \Cref{sec:discussionsystemparams}, many parameters are inferred near their prior bounds. \citet{Dorsman2026} analysed \ac{NICER} data of SAX J1808.4$-$3658 with a two hotspot model and accretion disc model, and found the same trend. When they used their \textsf{STU-Marg} model (which features non-pulsed component that is flexible but non-physical) \ac{m}, \ac{d}, \ac{i} moved away from the boundary edges. We disfavour the usage of the \textsf{STU-Marg} however, because it introduces an arbitrary choice of bounds to the non-pulsed component that shifted the inferred parameters for \citet{Dorsman2026}. Instead, we recommend a physics-based model for the disc emission or at least physics-based bounds if a phenomenological model is used.

\subsubsection{Improvements to the priors}\label{sec:discussionpriors}
Refining priors is another way of improving the accuracy and efficiency of analyses, meaning that parameter space is (accurately) narrowed down before analysis starts. As mentioned above, many parameters are inferred near their prior boundaries, which could be a hint that parameter space outside these prior distributions could lead to better fits to the data. In \Cref{sec:discussionsystemparams} we already mentioned that the priors for $i$ and $d$ were based on the distribution of values found so far in the literature, and these priors could be too narrow, so we recommend further exploration of these. 

\ac{m} and \ac{r} were also inferred to be close to their prior bounds in some cases. We used \ac{m} and \ac{r} prior bounds that were derived from an \ac{EoS}-informed prior, but not yet the actual multi-dimensional prior density distributions themselves. We recommend this for future work, as it would let one apply the best current prior knowledge of \ac{m} and \ac{r} from theory, experiment and other astrophysical data to the inference of \ac{AMP} properties \citep{HoogkamerRutherford2025}. As a final point, we also suggest narrowing the overall parameter space down further by self-consistently linking the stellar magnetic field and accretion rate to the inner disc radius and hotspot positions, shapes and sizes \citep[see e.g.][and references therein]{DiSalvo2022,Dorsman2025}.

\subsubsection{Future analysis of joint data}
\citet{Dorsman2025} and \citet{Salmi2025} both created synthetic data with the same parameters and used the data to test \ac{PPM} with a single instrument, but they did not yet test parameter recovery with the data sets jointly. In our analysis of the \ac{NICER} and \ac{IXPE} data, we found that results differ significantly depending on what combination method was applied. The technique of doing joint analysis thus deserves further testing. One natural recommendation is to test the combination methods with synthetic data to understand the trade-off between accuracy and computational cost.

We also recommend the combination of low energy, high energy, and polarized data,  because of the synergy of probing different physics in \acp{AMP}. For example, lower energy X-rays constrain the blackbody emission from the hotspots and accretion disc, whereas higher energy X-rays constrain the Comptonisation of radiation at the hotspots as well as reflection off the accretion disc (including the iron line at 6.4 keV). For the outburst of \SRGA{} studied in this paper, it might therefore be interesting to include the \ac{ART-XC} data in joint analysis. \ac{ART-XC} is sensitive at higher energy (up to 35 keV) compared to \ac{NICER} (2--10~keV). To add to this point, we inferred two different model fits for the plateau-like interpulse feature, but both are different compared to the proposed fit by \citet{Molkov2024discovery}. This warrants further investigation for example with \textit{ART-XC} data that are partially overlapping with \ac{NICER}. We note that one could also loop in \ac{IXPE}, {\it XMM-Newton}, {\it NuSTAR} and {\it HXMT} observations, which are (partially) overlapping as well. With regards to future observations, unfortunately, \ac{NICER} is currently non-active.\footnote{See update from August 5 2025 at \url{https://www.nasa.gov/missions/station/nicer-status-updates/}} However, \ac{eXTP} \citep{Zhang2025} and NewAthena \citep{Cruise2025} will be well-positioned to make joint observations, hosting wider band energy collectors and (in the case of \ac{eXTP}) polarimetric collectors on board, both with much larger effective areas compared to their predecessors.

\section{Conclusion}\label{sec:conclusion}
We fitted the \ac{NICER} and \ac{IXPE} pulse profiles of the 2024 outburst \SRGA{} with a \ac{NS} and accretion disc model, for the first time including polarization data. We analysed the two data sets first separately and then jointly, which was done in three different ways to verify robustness. First, we combined the posterior distributions that resulted from the separate analysis. Second, we inserted posterior distributions of constant (i.e. shared) parameters resulting in one analysis as priors into the next. That was done in both directions to test robustness. Third, in \Cref{sec:appendix_sts}, we used a joint model to simultaneously fit both data sets. Due to high computational cost, simplified identical antipodal hotspots were adopted there, which unfortunately proved insufficient to fit the data.

We consistently inferred a high \ac{NS} mass at 1.9$-$2.2\ M$_\odot$ but found two possible modes for the radius, either at 9.7$-$13.1~km or at 13.6$-$14~km. For these results it is important that we note that both the mass and radius inferences may not be robust because they touch the upper bound of their priors (2.2 M$_\odot$ and 14 km respectively), and we recommended further analyses with other priors and improved models. Furthermore, we inferred a \ac{NS} with a cooler, large primary hotspot that is centred close to the rotational axis, which produces the bulk of the non-pulsed radiation. The secondary hotspot is hotter and is the main contributor to the pulse profile shape.

Because many of the inferred parameters were near the edges of their respective priors, we are concerned that some physics is missing in the model (such as disc reflection or reprocessing), which could cause parameter inferences to be biased. To examine such a potential bias we recommend reanalysis with improved models and with more conservative or physics-informed priors where possible.

There were notable shifts in the inferred parameters between the combination methods. Therefore the results are not fully robust, and we recommend testing with synthetic data to better understand the trade-off between accuracy and computational cost of each method. We also recommend improving the computational efficiency of the analysis and considering only simultaneously observed data to make the third combination method computationally viable. We expect that the combination of low energy, high energy, and polarized data will lead to complementary constraints because these probe different physics. 

\section*{Acknowledgements}
B.D. and A.L.W. acknowledge support from ERC Consolidator grant No. 865768 AEONS (PI: Watts). T.S. acknowledges funding by the Research Council of Finland grant No.~368807. M.N. is a Fonds de Recherche du Quebec– Nature et Technologies (FRQNT) postdoctoral fellow. 
T.S., A.B., and J.P. were supported by the Research Council of Finland Centre of Excellence in Neutron-Star Physics (grants 374063 and 374064).
M.H., Y.K. and A.L.W. acknowledge support from the NWO grant ENW-XL OCENW.XL21.XL21.038 \textit{Probing the phase diagram of Quantum Chromodynamics} (PI: Watts). S.G. acknowledges the support of the CNES and of the Agence Nationale pour la Recherche through grant ANR-25-CE31-7901-01 (ANR DENSER). C.M. and A.P. acknowledge support from INAF (Research Grant FANS and PULSE-X), the Italian Ministry of University and Research (PRIN MUR 2020, Grant 2020BRP57Z, GEMS), and Fondazione Cariplo/Cassa Depositi e Prestiti (Grant 2023-2560).
This work was supported in part by NASA through the NICER mission. We acknowledge NWO for providing access to Snellius, hosted by SURF through the Computing Time on National Computer Facilities call for proposals.
This work used the Dutch national e-infrastructure with the support of the SURF Cooperative using grant no. EINF-11691 and is subsidized by NWO Domain Science. 
This research was supported by the International Space Science Institute (ISSI) in Bern, through International Team project 25-657 'Polarimetric Insights into Extreme Magnetism’. We acknowledge extensive use of NASA’s Astrophysics Data System (ADS) Bibliographic Services and the ArXiv.

\section*{Data Availability}
A full reproduction package is available at \href{https://doi.org/10.5281/zenodo.18376256}{10.5281/zenodo.18376256}

\bibliographystyle{mnras}
\bibliography{bibliography}

\appendix
\section{Combination method 3 (simultaneous fit)}\label{sec:appendix_sts}
This appendix presents the results of the analysis using a joint model to simultaneously fit the \ac{NICER} and \ac{IXPE} data. Because the observations of the data sets were disjoint, two instances of all variable parameters are used. For some parameters, that may turn out unnecessary if we end up inferring similar values. However, we choose to always treat the two instances as independent parameters because we do not know a priori how different the inferred values will be. For this reason -- and because the emission is computed twice (for \ac{NICER} and for \ac{IXPE}) -- the analysis has significantly higher computational cost. To get a run to complete within the budget, we reduced the model complexity from \stu{} to \sts{} (defined in \Cref{sec:model}). This brings the total parameters down from a potential 35 to 23. Still, at 115 \ac{kch}, this run was significantly more expensive than even the \ac{IXPE} and \ac{NICER} runs added together, as shown in \Cref{tab:inferences,tab:inferences_appendix}.

When we look at the inferred parameters, we note that the inferred \ac{m} is now towards the lower bound (rather than upper bound) of the prior, at $M=1.015^{+0.024}_{-0.011}$M$_\odot$. On the other hand, the inferred \ac{r} is consistent with high radius results preferred by the \stu{} results that analyse \ac{NICER} data first, at $13.90^{+0.08}_{-0.16}$ km. \ac{d} and \ac{i} tend towards the lower bound of their priors, which is a similar finding to all previous results. \ac{nh}, at $(25.52 \pm 0.06) \times 10^{21}$ cm$^{-2}$, is roughly in the middle of the inferred values from the \ac{NICER} and \ac{IXPE} analyses. We also find again that the decrease in flux is explained mainly by a decrease of $T_{\rm seed,p}$, consistent with the idea that the accretion rate decreases.

It is also interesting to note that the inner disc radius inferred for the \ac{NICER} data is very small in this analysis, much smaller compared to the \stu{} findings, and blocks almost the entire southern hemisphere and secondary hotspot. This finding is not consistent with the classic \ac{AMP} picture where accreted material from the inner accretion disc that truncates some way from the stellar surface is funnelled towards the magnetic poles of the \ac{NS}, forming hotspots, but instead looks more like boundary layer accretion. The hotspots are inferred to be large during both the \ac{NICER} and \ac{IXPE} observations, but the spot colatitudes have increased significantly between the two observations.

\begin{figure}
    \centering
    \includegraphics[width=\linewidth]{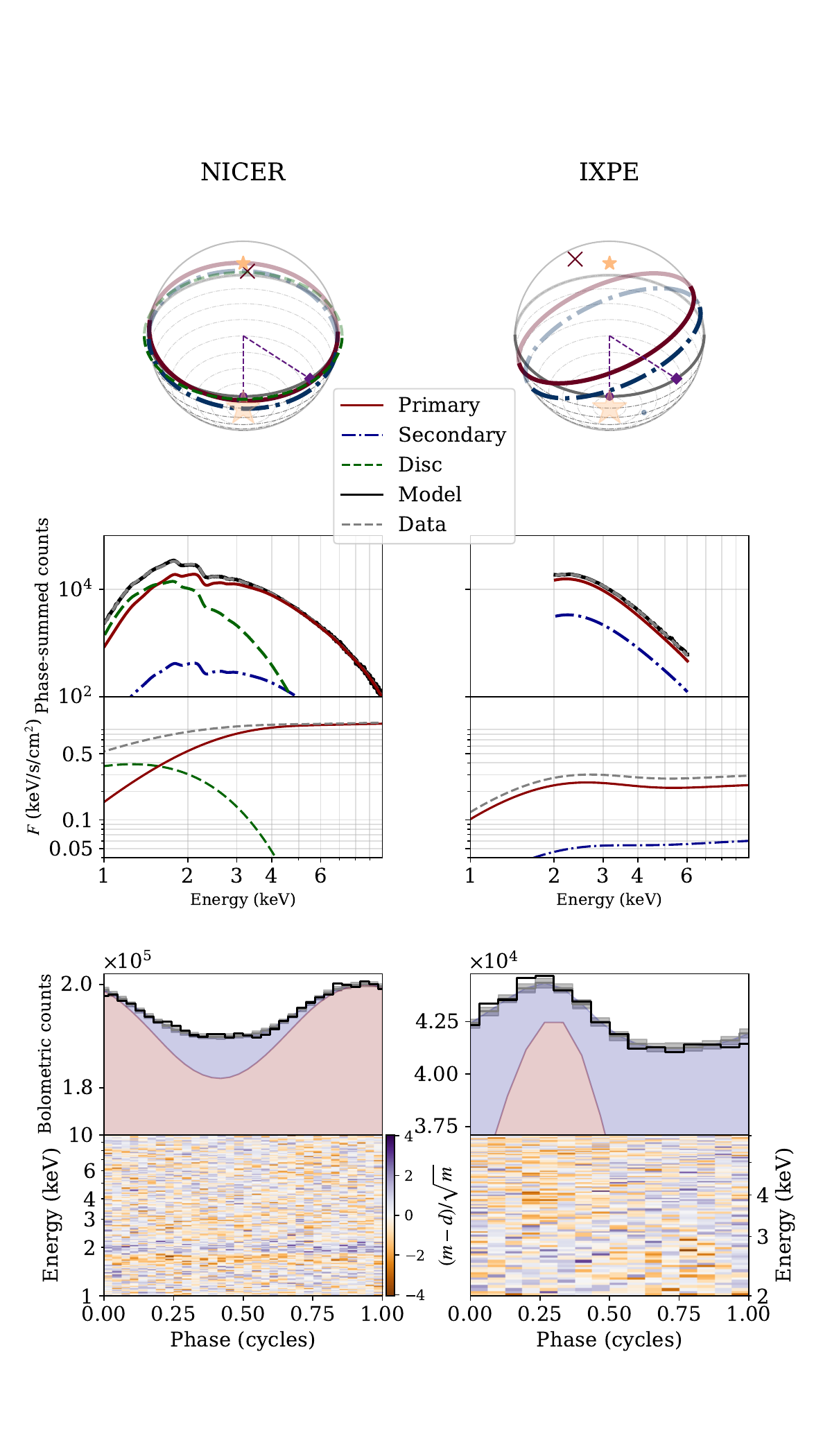}
    \caption{Same figure as \Cref{fig:diagnostics}, but for the joint model analysis. In this case, the left panels show the projection, spectra, pulse profiles and residuals associated with the model during the \ac{NICER} observation, and the right panels with the \ac{IXPE} (DU$_1$) observation. DU$_2$ and DU$_3$ are not shown, but they are very similar to DU$_1$. See the caption of \Cref{fig:diagnostics} for more detail on all the components in the plot.}
    \label{fig:diagnostics_STS}
\end{figure}

When we look at the fit statistics, we see that the joint model fits of either data set are significantly worse than the fits in the main body of the paper. The $\chi^2/$DOF for the \ac{NICER} data is $30772/28777 = 1.0693$, corresponding to a p-value of $2.0629\times10^{-16}$, and the $\chi^2/$DOF for the Stokes $I$ data of the three \ac{IXPE} detectors is $5250/4825 = 1.0881$, corresponding to a p-value of $1.2619\times10^{-5}$. We see again the \ac{NICER} aspects in the \ac{NICER} residual. In addition, here the worse fit is manifested in a phase-dependent structure visible in both residuals, shown by \Cref{fig:diagnostics_STS}. This indicates that the \sts{} joint model is strongly disfavoured compared to the separate \stu{} analyses. One caveat is that there were only 1000 live points used here, and that can be considered insufficient for an analysis with 23 parameters, especially if the likelihood surface is complex. Due to the expected high computational cost, these results were not tested for convergence against changes to the sampler settings. Another caveat in this analysis is that four parameters: \ac{m}, \ac{d}, $T_{\rm e,N}$, and $\alpha_{\rm N}$ all converged towards prior boundaries, hinting that parameter space outside current prior boundaries could lead to better fits. 

To conclude, the antipodal and identical hotspots that \sts{} enforces, appear a serious restriction that limits the ability of the model to fit the data. For future work, we recommend the application of the \stu{} configuration instead. We describe methods to overcome the computational cost in \Cref{sec:discussion_speedup}.

\begin{table}
\centering
\caption{An overview of the analysis with the joint model and the inferred parameters. \label{tab:inferences_appendix} This table follows the format of \Cref{tab:inferences}, see also the caption of that table.}
\begin{tabular}{lc}
\hline
\hline
Name & \textbf{Combination method 3} \T\B\\
\hline
Data & NICER, \ac{IXPE} $IQU$\\
Hotspot config. &\sts\\
\textbf{Sampling aspects} &\T\B\\
Num. params. & 23\\
Samp. eff. (X-PSI) & 0.1\\
Run cost (kch) & 115.2 \\
Live points & 1000\\
$\ln(\mathcal{Z})$ & $-$1.5955e+5\\
\textbf{\ac{NICER} fit quality} & \T\B\\
$\chi^2$/DOF & 30744/28777\\
$\hat{\chi}^2$ & 1.0683\\
P-value & 5.268e$-$16\\
\textbf{\ac{IXPE} fit quality} & \T\B\\
$\chi^2$/DOF & 5251/4825\\
$\hat{\chi}^2$ & 1.0888\\
P-value & 1.1861e-5\\
\hline
$M\;\mathrm{[M}_{\odot}\mathrm{]}$ & $1.015_{-0.011}^{+0.024}$ \T\B \\
$R_{\mathrm{eq}}\;\mathrm{[km]}$ & $13.90_{-0.16}^{+0.08}$ \T\B \\
$D \;\mathrm{[kpc]}$ & $3.13_{-0.09}^{+0.09}$ \T\B \\
$i\;\mathrm{[deg]}$ & $50.25_{-0.19}^{+0.41}$ \T\B \\
$N_\mathrm{H}\;[10^{21} \mathrm{cm}^{-2}]$ & $25.52_{-0.06}^{+0.06}$ \T\B \\
$\phi_\mathrm{p,N}\;\mathrm{[cycles]}$ & $0.056_{-0.003}^{+0.003}$ \T\B \\
$\theta_\mathrm{p,N}\;\mathrm{[deg]}$ & $6.9_{-0.8}^{+1.0}$ \T\B \\
$\zeta_\mathrm{p,N}\;\mathrm{[deg]}$ & $86.42_{-0.21}^{+0.17}$ \T\B \\
$T_\mathrm{seed,p,N}\;\mathrm{[keV]}$ & $0.812_{-0.013}^{+0.012}$ \T\B \\
$T_\mathrm{e,p,N}\;\mathrm{[keV]}$ & $21.3_{-0.6}^{+0.7}$ \T\B \\
$\tau_\mathrm{p,N}\;[-]$ & $2.40_{-0.04}^{+0.03}$ \T\B \\
$\phi_\mathrm{p,I}\;\mathrm{[cycles]}$ & $-0.319_{-0.004}^{+0.004}$ \T\B \\
$\theta_\mathrm{p,I}\;\mathrm{[deg]}$ & $24.0_{-2.5}^{+2.5}$ \T\B \\
$\zeta_\mathrm{p,I}\;\mathrm{[deg]}$ & $84.2_{-0.5}^{+0.4}$ \T\B \\
$T_\mathrm{seed,p,I}\;\mathrm{[keV]}$ & $0.568_{-0.007}^{+0.007}$ \T\B \\
$T_\mathrm{e,p,I}\;\mathrm{[keV]}$ & $46.2_{-4.5}^{+5.3}$ \T\B \\
$\tau_\mathrm{p,I}\;[-]$ & $1.30_{-0.05}^{+0.06}$ \T\B \\
$T_\mathrm{in,N}\;\mathrm{[keV]}$ & $0.54_{-0.01}^{+0.01}$ \T\B \\
$R_\mathrm{in,N}\;\mathrm{[km]}$ & $14.46_{-0.17}^{+0.13}$ \T\B \\
$\alpha_\mathrm{NICER}\;[-]$ & $0.811_{-0.009}^{+0.019}$ \T\B \\
$\alpha_\mathrm{DU2}\;[-]$ & $1.0218_{-0.0018}^{+0.0018}$ \T\B \\
$\alpha_\mathrm{DU3}\;[-]$ & $0.9939_{-0.0017}^{+0.0017}$ \T\B \\
$\chi_0\;\mathrm{[rad]}$ & $-0.7_{-0.6}^{+1.8}$ \T\B \\
\hline
\end{tabular}
\end{table}

\begin{figure}
    \centering
    \includegraphics[width=\linewidth]{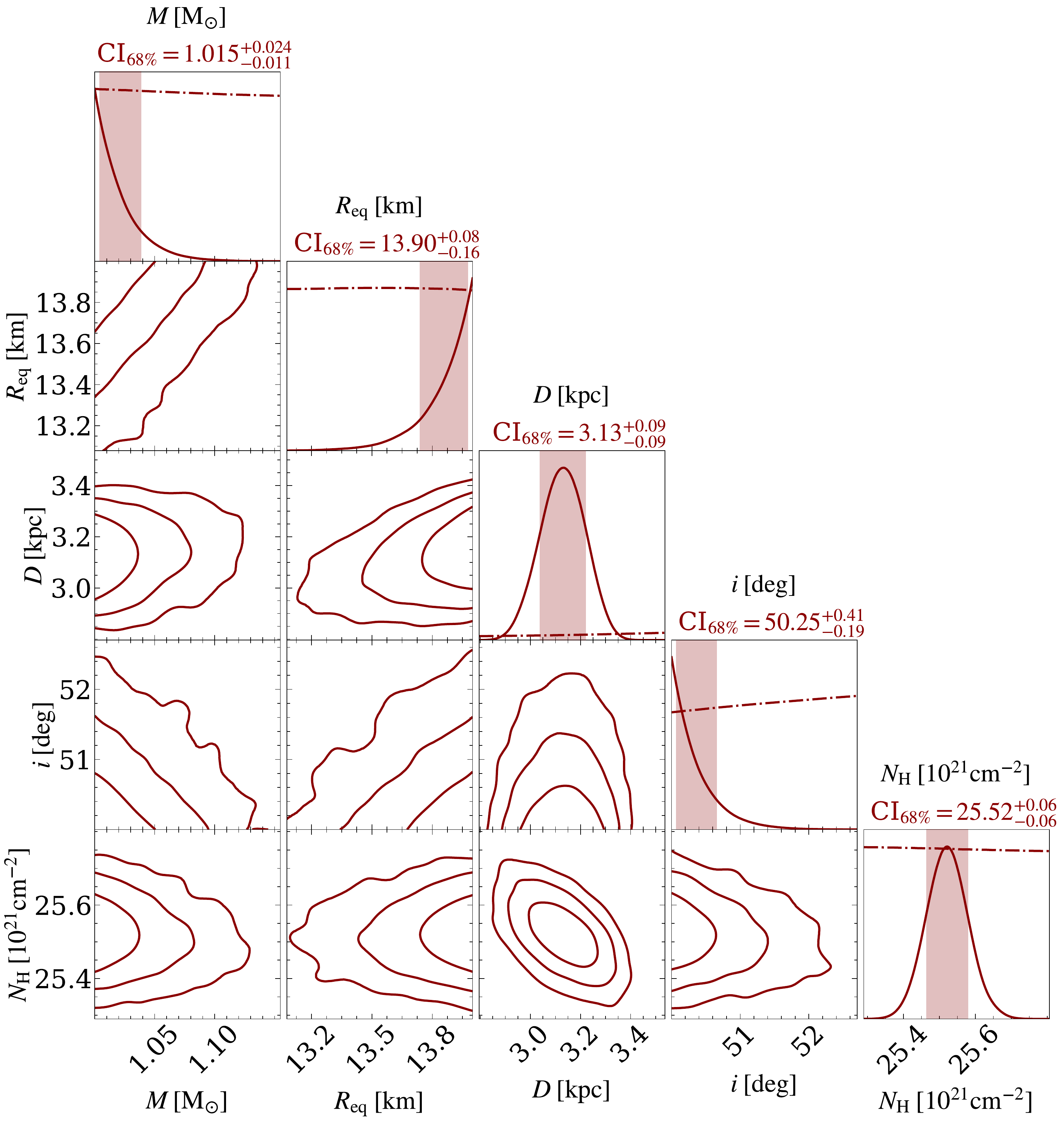}
    \caption{Posterior distributions of the inferred constant parameters of the joint model analysis (combination method 3, see \Cref{sec:appendix_sts}). See the caption of \Cref{fig:cornerplot_kde_combination} for a detailed description of the plot elements.}
    \label{fig:placeholder}
\end{figure}

\section{Extra corner plots}\label{sec:appendix_extra_corner_plot}
\begin{figure*}
    \centering    
    \includegraphics[width=\textwidth]{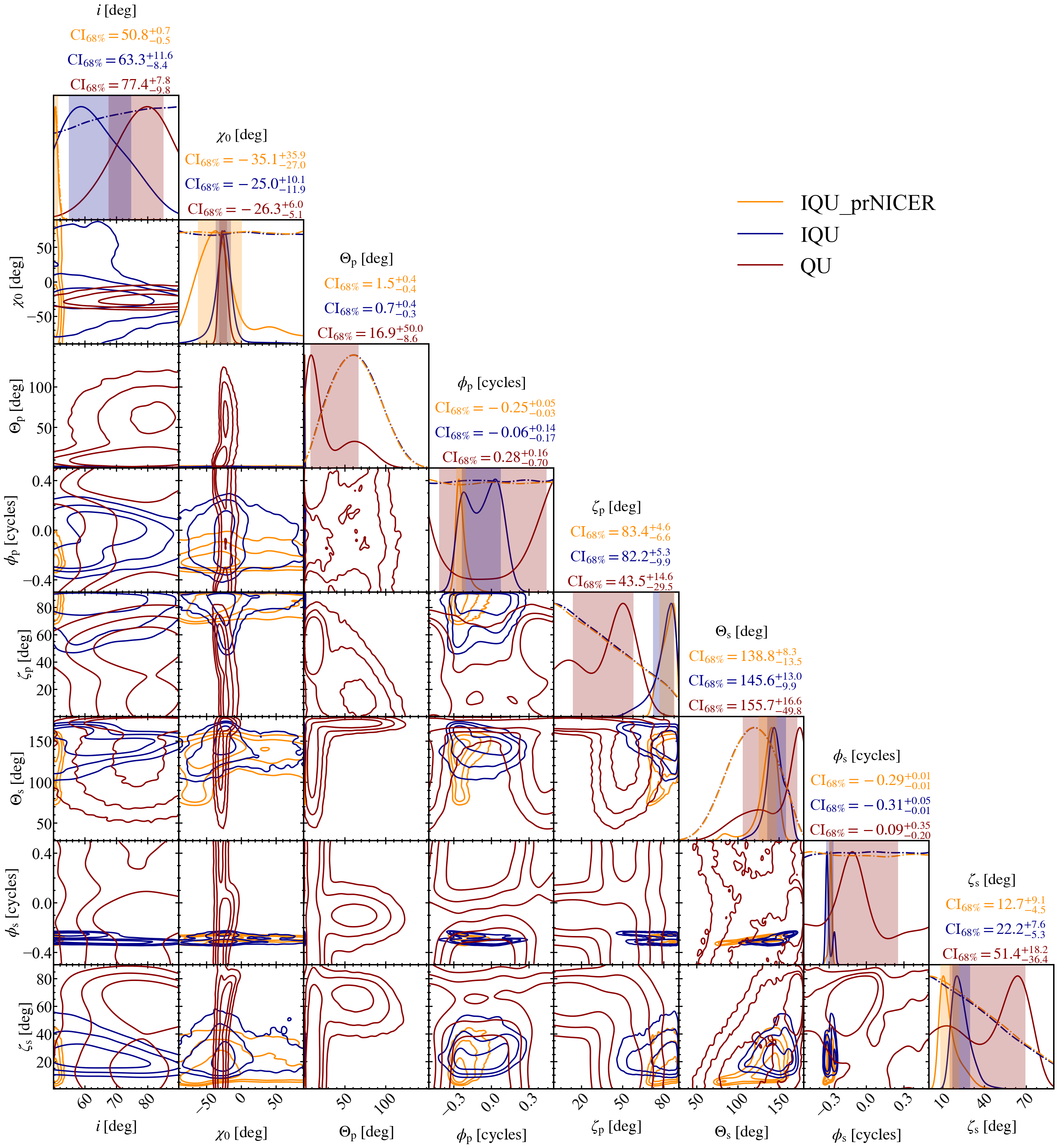}
    \caption{Posterior distributions of the geometry parameters for the \ac{IXPE} analyses.
     See the parameter and model definitions in \Cref{tab:parameters}, \Cref{sec:results_ixpe}, and \Cref{sec:results_ixpe_nicer-informed}.
     The dash–dotted curves represent the prior distributions (which are identical for the IQU and QU cases).
     The 1D shaded intervals contain 68.3~per cent of the posterior mass, and the 2D contours contain 68.3, 95.4, and 99.7~per cent of the posterior mass.
     The IQU posteriors of $\Theta_{\rm p}$ are difficult to see, because they are at the lower bound of the prior.
    }    
    \label{fig:cornerplot_IXPEonly_pol}
\end{figure*}

\begin{figure*}
    \centering
    \includegraphics[width=\textwidth]{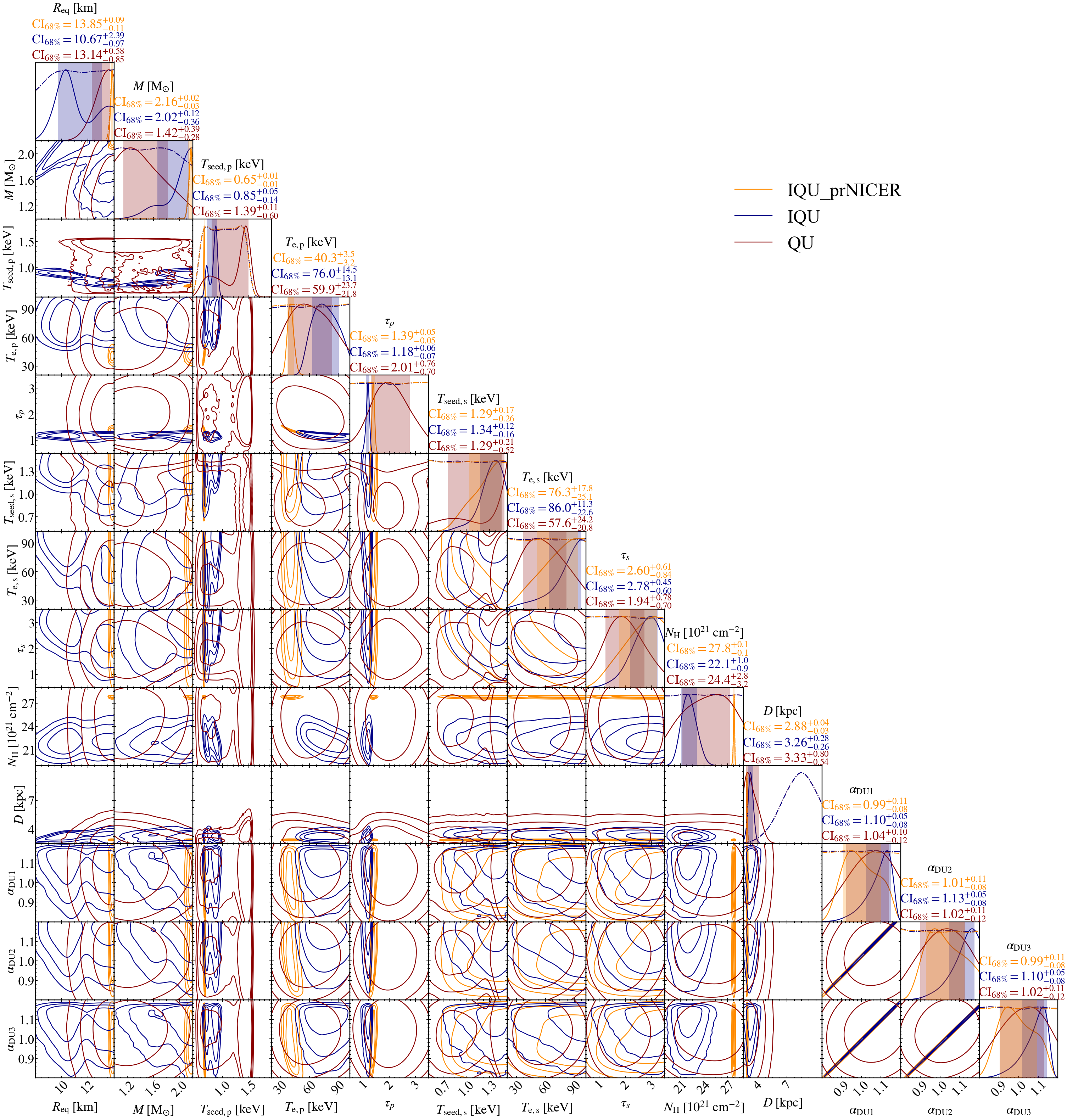}
    \caption{Posterior distributions of the non-geometry parameters for the \ac{IXPE} analyses.
     See the parameter and model definitions in \Cref{tab:parameters}, \Cref{sec:results_ixpe}, and \Cref{sec:results_ixpe_nicer-informed}.
     The dash–dotted curves represent the prior distributions (which are identical for the IQU and QU cases).
     The 1D shaded intervals contain 68.3~per cent of the posterior mass, and the 2D contours contain 68.3, 95.4, and 99.7~per cent of the posterior mass.
    }
    \label{fig:cornerplot_IXPEonly_rest}
\end{figure*}

\begin{figure*}
    \centering
    \includegraphics[width=\textwidth]{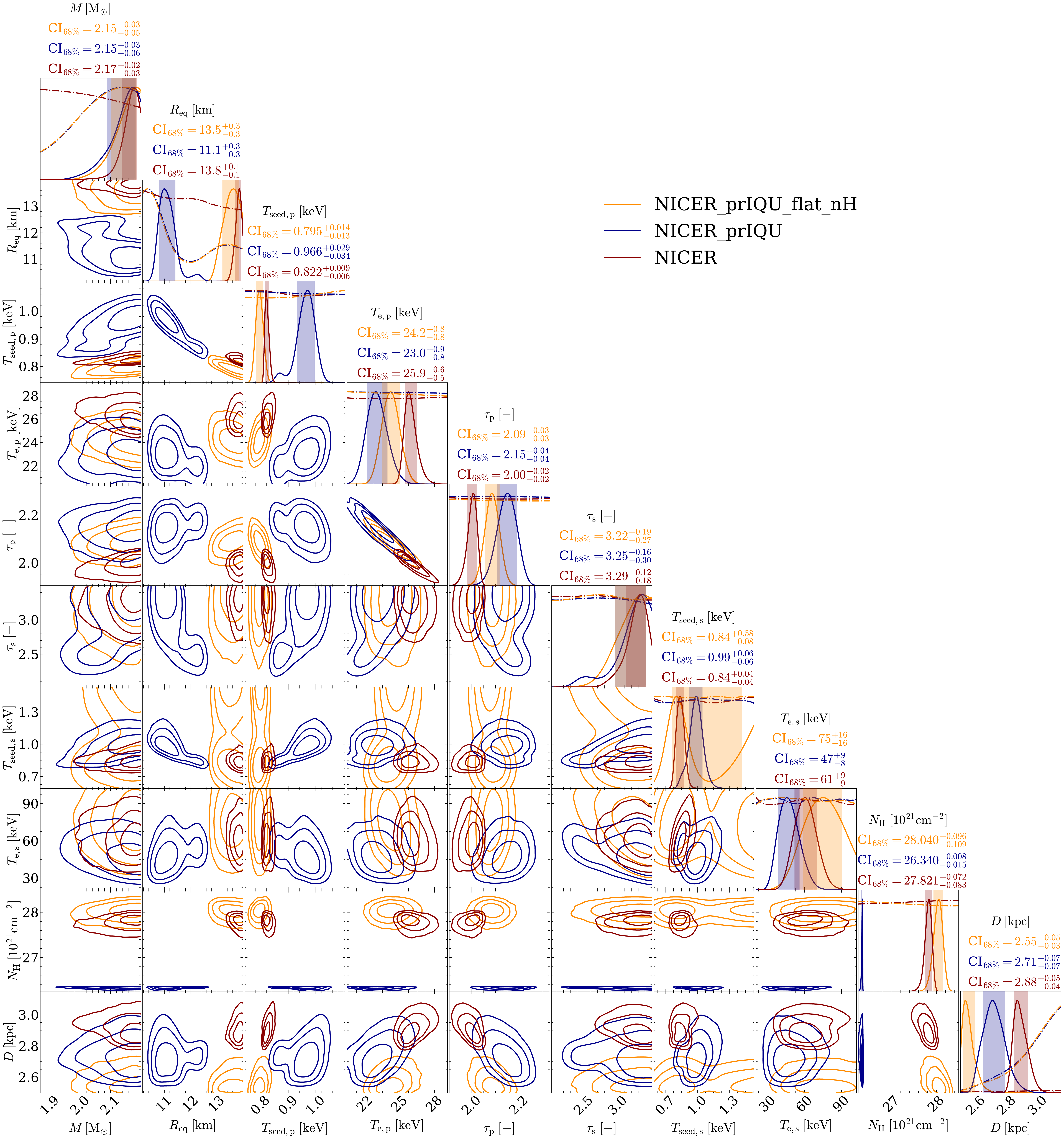}
    \caption{Posterior distributions of the non-geometry parameters for the \ac{NICER} analyses. See the parameter and model definitions in \Cref{tab:parameters}, \Cref{sec:model}. The dash–dotted curves represent the prior distributions (which only differ in \ac{nh}). The 1D shaded intervals contain 68.3~per cent of the posterior mass, and the 2D contours contain 68.3, 95.4, and 99.7~per cent of the posterior mass.}
    \label{fig:cornerplot_NICERonly_flux}
\end{figure*}

\begin{figure*}
    \centering
    \includegraphics[width=\textwidth]{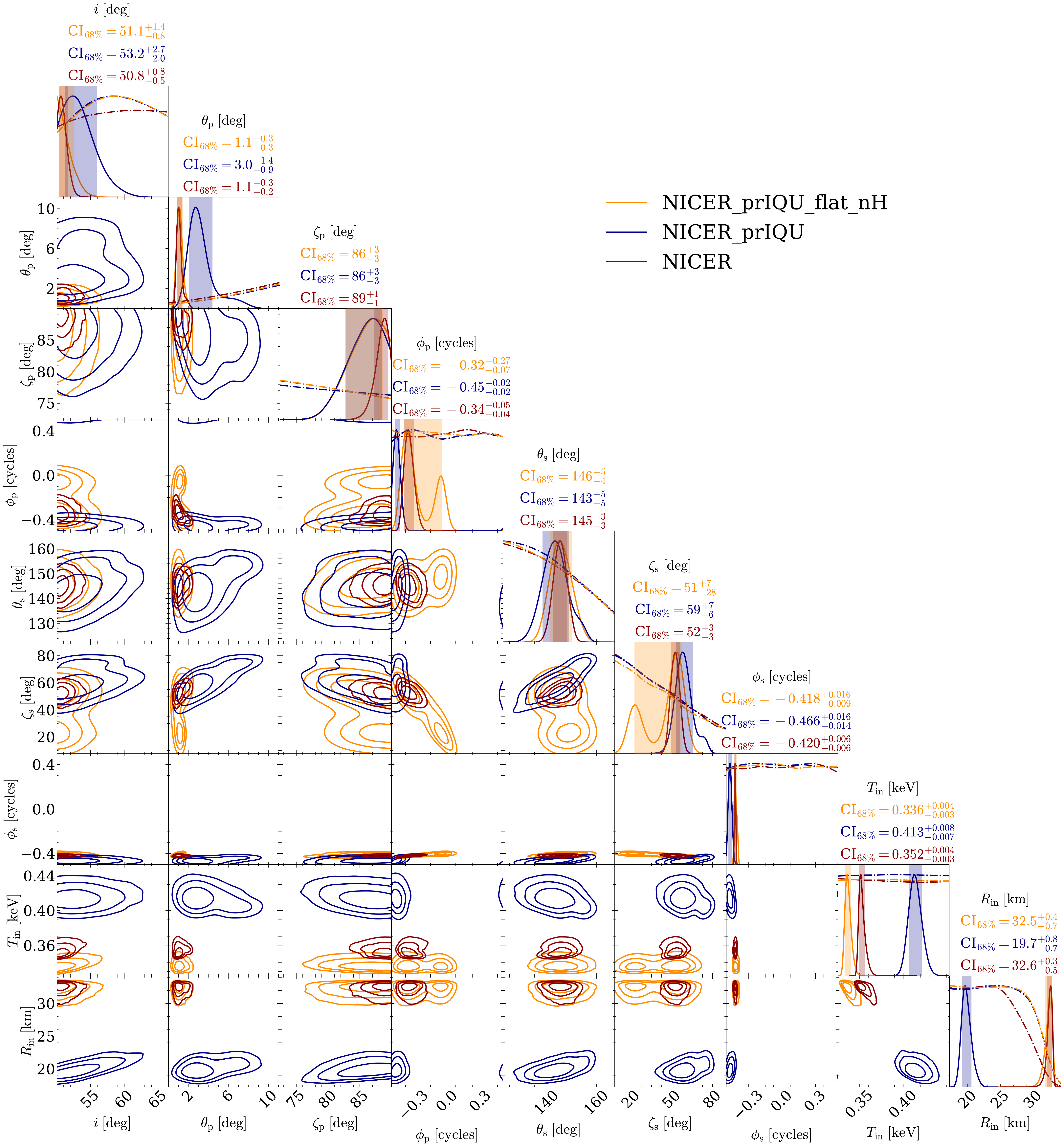}
    \caption{Posterior distributions of the geometry parameters for the \ac{NICER} analyses. See the parameter and model definitions in \Cref{tab:parameters}, \Cref{sec:model}. The dash–dotted curves represent the prior distributions (which only differ in \ac{nh}). The 1D shaded intervals contain 68.3~per cent of the posterior mass, and the 2D contours contain 68.3, 95.4, and 99.7~per cent of the posterior mass.}  \label{fig:cornerplot_NICERonly_geom}
\end{figure*}

\begin{figure*}
    \centering
    \includegraphics[width=\textwidth]{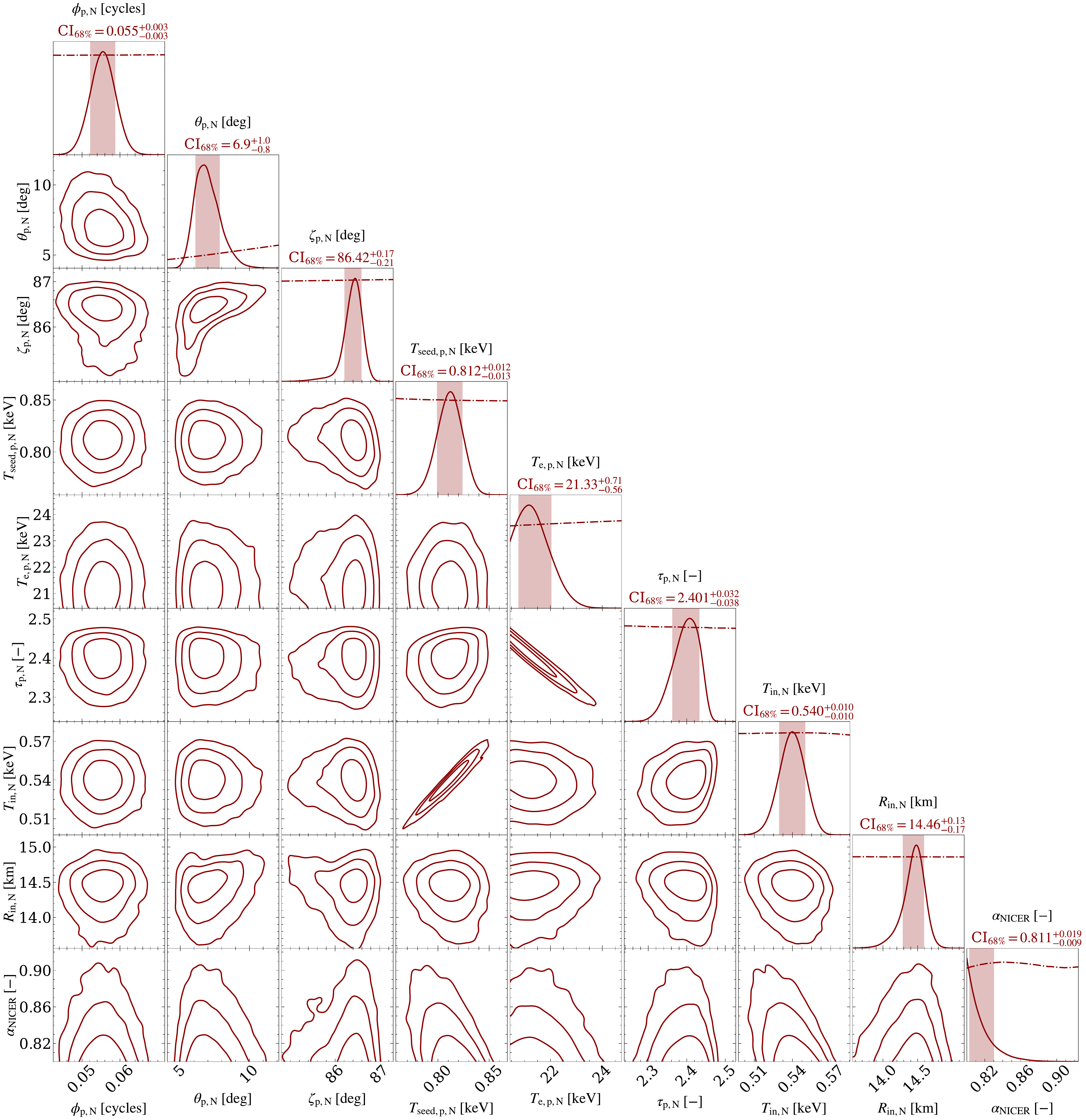}
    \caption{Posterior distributions of the variable parameters during the \ac{NICER} observation for the joint model analysis. See the parameter and model definitions in \Cref{tab:parameters}, \Cref{sec:model}. The dash–dotted curves represent the prior distributions (which only differ in \ac{nh}). The 1D shaded intervals contain 68.3~per cent of the posterior mass, and the 2D contours contain 68.3, 95.4, and 99.7~per cent of the posterior mass.}  \label{fig:cornerplot_NICER_STS_var}
\end{figure*}

\begin{figure*}
    \centering
    \includegraphics[width=\textwidth]{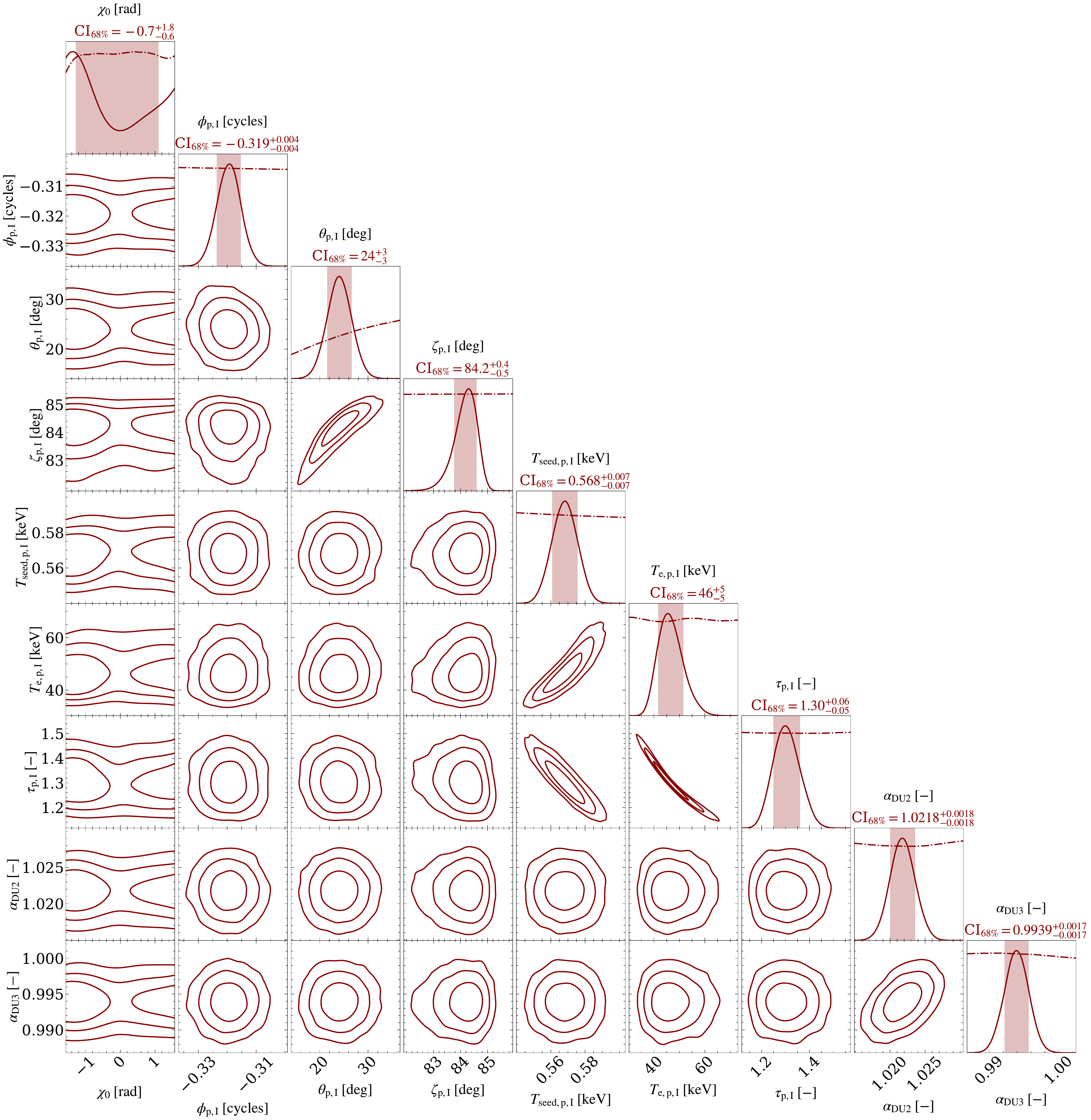}
    \caption{Posterior distributions of the variable parameters during the \ac{IXPE} observation for the joint model analysis. See the parameter and model definitions in \Cref{tab:parameters}, \Cref{sec:model}. The dash–dotted curves represent the prior distributions (which only differ in \ac{nh}). The 1D shaded intervals contain 68.3~per cent of the posterior mass, and the 2D contours contain 68.3, 95.4, and 99.7~per cent of the posterior mass.}  \label{fig:cornerplot_IXPE_STS_var}
\end{figure*}

\Cref{fig:cornerplot_IXPEonly_pol,fig:cornerplot_IXPEonly_rest,fig:cornerplot_NICERonly_flux,fig:cornerplot_NICERonly_geom} show the full posterior distributions of all the \stu{} analyses included in this work. This includes the \ac{IXPE} Stokes $Q$ and $U$ analysis, in which the Stokes $I$ data was not used. \Cref{fig:cornerplot_NICER_STS_var,fig:cornerplot_IXPE_STS_var} show the posterior distributions of the variable parameters (as observed by \ac{NICER} or \ac{IXPE}) resulting from combination method 3.

\bsp	
\label{lastpage}
\end{document}